\newif\ifShowKeys
\ShowKeysfalse

\documentclass[11pt,a4paper]{article} 
\pdfoutput=1
\usepackage[no-natbib-sort]{my-jheppub}

\usepackage{amsmath, amssymb}

\usepackage{bm}
\usepackage{environ}
\usepackage{mathrsfs}
\usepackage{array,arydshln}

\usepackage{graphicx,epsfig}
\usepackage{epic}
\usepackage{youngtab}
\usepackage{float}
\usepackage{color}
\definecolor{maroon}{rgb}{0.8,0.3,0.}

\usepackage{slashed}
\usepackage[nodayofweek]{date time}

\ifShowKeys \usepackage{showkeys} \fi

\usepackage{hyperref}

\usepackage{aurical}
\usepackage[T1]{fontenc}

\allowdisplaybreaks

\newcommand{\be}{\begin{equation}}
\newcommand{\ee}{\end{equation}}

\newcommand{\mc}{\mathcal }

\newcommand{\bv}{\begin{verbatim}}
\newcommand{\ev}{\end{verbatim}}

\newcommand{\la}{\label}
\newcommand{\Vb}{\overline V}

\title{The ground state of long-range  Schr\"odinger equations 
 and static $q\overline{q}$ potential}
\author[a,b]{Matteo Beccaria} 
\author[a]{, Giorgio Metafune} 
\author[a,b]{, Diego Pallara} 

\abstract{
Motivated by the recent results in 
\href{http://arxiv.org/abs/1601.05679}{arXiv:1601.05679} about the quark-antiquark potential
in $\mc N=4$ SYM, we reconsider the problem of computing the asymptotic weak-coupling expansion 
of the ground state energy of a certain class of 1d Schr\"odinger operators 
$-\frac{d^{2}}{dx^{2}}+\lambda\,V(x)$ with long-range
potential $V(x)$. In particular, we  consider even potentials obeying $\int_{\mathbb R}dx\, V(x)<0$
with large $x$ asymptotics 
$V\sim -a/x^{2}-b/x^{3}+\cdots$. The  associated Schr\"odinger operator is known to admit a bound state for 
$\lambda\to 0^{+}$, but the binding energy is rigorously  non-analytic at $\lambda=0$.
Its asymptotic expansion starts at order $\mc O(\lambda)$, but contains higher corrections 
 $\lambda^{n}\,\log^{m}\lambda$ with 
all $0\le m\le n-1$ and 
standard Rayleigh-Schr\"odinger perturbation theory fails order by order in $\lambda$.
We discuss various analytical tools to tame this problem and provide the general expansion of the 
binding energy at $\mc O(\lambda^{3})$
in terms of  quadratures. 
The method is tested on a soluble potential that is fully under control, and on 
various  non-soluble cases as well. A supersymmetric case, arising in the study of the quark-antiquark 
potential in  $\mc N=6$ ABJ(M) theory, is  also 
exploited to provide a further 
non-trivial consistency check.
Our analytical results confirm at third order a remarkable 
exponentiation  of the leading infrared logarithms, first noticed in  $\mc N=4$ SYM 
where it may be proved by Renormalization Group arguments.
We prove this interesting feature at all orders at the level of the Schr\"odinger equation 
for general potentials in the considered class.
\vfill }

\affiliation[a]{Dipartimento di Matematica e Fisica Ennio De Giorgi,\\
Universit\`a del Salento, Via Arnesano, 73100 Lecce, 
Italy} 

\affiliation[b]{INFN, Via Arnesano, 73100 Lecce, Italy}

\emailAdd{matteo.beccaria@le.infn.it} 
\emailAdd{giorgio.metafune@unisalento.it} 
\emailAdd{diego.pallara@unisalento.it} 

\begin{document}



\maketitle
\flushbottom

\section{Introduction}

Let us consider the Schr\"odinger operator in one dimension
\be
\label{1.1}
H = -\frac{d^{2}}{dx^{2}}+\lambda\,V(x),\qquad x\in \mathbb{R}, \ \lambda>0.
\ee
A  variational argument shows that if $V(x)\le 0$, with $V(x)<0$ on an open set, and $V(x)\to 0$
at infinity, then $H$ has a bound state for all $\lambda>0$. The lowest eigenvalue $E_{0}(\lambda)$
is real analytic for $\lambda>0$ under mild conditions on $V$ \cite{reed1978analysis}.
Interesting questions are analyticity at $\lambda=0$ and the existence of bound states for small $\lambda$
when $V$ is somewhere positive. \footnote{The large $\lambda$ limit is completely different, see for instance 
\cite{MR0246593,MR0247837}, although in some cases, scale invariance connects it to the weak coupling 
regime
\cite{avron1979strongly}. 
}
The analysis of \cite{MR0404846} showed that when 
$\int_{\mathbb R}dx\,V(x)\le 0$ and $\int_{\mathbb R}dx\,(1+x^{2})\,|V(x)| < \infty$, there is a unique bound state
for small $\lambda$. The ground state (binding) energy may be estimated in this case by the formula 
\be
\label{1.2}
\sqrt{-E_{0}(\lambda)} = -\frac{1}{2}\,\lambda\,\int_{\mathbb R}dx\,V(x)
-\frac{1}{4}\,\lambda^{2}\,\int_{\mathbb R}dx\,dy\,|x-y|\,V(x)\,V(y)+\mc O(\lambda^{2}).
\ee
If for some  $a>0$ we have $\int_{\mathbb R} dx\, e^{a\,|x|}\,V(x)<\infty$, then
the quantity $\sqrt{-E_{0}(\lambda)}$
is analytic at $\lambda=0$. \footnote{This short-range case is much simpler \cite{MR610664}
 and may be treated by standard
perturbation theory. For instance, a quite compact extension of (\ref{1.2}) to order $\mc O(\lambda^{4})$
is discussed in  \cite{patil1980t,gat1993new,Collins:1995hd} and may be easily extended at higher orders.
The first term
of (\ref{1.2}) may be found immediately by applying the Feynman-Hellman theorem \cite{Feynman:1939zza}.
} If the potential is such that 
$\int_{\mathbb R}dx\,(1+x^{2})\,|V(x)|  = \infty$, and $V(x) \sim -a\,|x|^{-\beta}$ at infinity, the results of 
 \cite{Blankenbecler:1977pf} imply the following. For $2<\beta<3$, the estimate (\ref{1.2}) is still valid.
 For $\beta=2$, there is a unique bound state for $\lambda\to 0^{+}$ provided that $\int_{\mathbb R}dx\,V(x)\le 0$.
 However, (\ref{1.2}) is violated because the r.h.s. develops a term $\lambda^{2}\,\log\lambda$.
 If $1<\beta<2$, there are infinitely many bound states for any $\lambda>0$
 and (\ref{1.2}) is valid when restricted to the first term only, {\em i.e.}
 \be
\label{1.3}
\sqrt{-E_{0}(\lambda)} = -\frac{1}{2}\,\lambda\,\int_{\mathbb R}dx\,V(x)+\mc O(\lambda).
\ee
Here, we shall be concerned with the $\beta=2$ case. Further discussion of other long-range cases 
may be found in \cite{avron1981,klaus1979remark} and in the review \cite{MR1768631}.
Thus, in this paper, we shall address the problem (\ref{1.1}) with a potential in the class defined by 
the conditions
\be
\label{1.4}
V(x) \stackrel{x\to \infty}{\sim} -a\,x^{-2}-b\,|x|^{-3}+\cdots,\qquad \int_{\mathbb R}dx\, V(x) < 0,
\ee
and, for simplicity, we shall also assume it to be even $V(x) = V(-x)$. The modified form of (\ref{1.2}) under the
conditions in (\ref{1.4}) has been found in \cite{Blankenbecler:1977pf} and reads
\be
\label{1.5}
\sqrt{-E_{0}(\lambda)} = -\bigg[\frac{1}{2}\,\lambda+a\,\lambda^{2}\,\log\lambda\bigg]\,\int_{\mathbb R} dx\, V(x) + \mc O(\lambda^{2}).
\ee
A non-analytic logarithmic factor enhances the second order term. Standard 
Rayleigh-Schr\"odinger perturbation theory completely fails and this term is infinite. \footnote{
It is important to emphasize that we are not dealing with the divergence of the perturbative series due to 
a vanishing radius of convergence. The long-distance behaviour of the potential is such that the single terms
in the Rayleigh-Schr\"odinger perturbative expansion are separately divergent.
}
In general,
it is very difficult to recover results like (\ref{1.5}) from some kind of regularized perturbation theory.
A very illustrative example are crude infrared cutoffs like $|x|\le L$. \footnote{
The study of the infrared problems that appear when $V(x)$ is treated as a perturbation of the  laplacian dates back to the 
initial developments of quantum mechanics. A celebrated example is E. P. Wigner's discussion of 
the hydrogen atom in the classical paper \cite{Wigner}, see also \cite{trees1956application}. }
Finite size effects may be suppressed at 
large  $L$ but by factors like $e^{-\lambda\,g}$ or similar. Expanding first in $\lambda$ shows that the cutoff dependence
is increasingly bad at higher orders.

A modern application of results like (\ref{1.5}) is to the study of the quark-antiquark static potential in flat space
for $\mc N=4$ super Yang-Mills theory with gauge group $SU(N_{c})$.~\footnote{Another (older) interesting physical application of (\ref{1.1}) with potential in the class (\ref{1.4})
occurs in the study of  wetting transition
of 2d surfaces  \cite{kroll1983universality}. In that context, 
the interface free-energy density is given in the thermodynamic limit by the lowest eigenvalue $E_{0}$
of (\ref{1.1}) where $x$ denotes the perpendicular distance of the interface from the substrate, and
$V(x)$ is the potential well that localizes the interface below the transition temperature, related to $\lambda$.
} 
The  static potential is extracted from a pair of anti-parallel Wilson lines separated by the distance $r$ 
\cite{Erickson:1999qv,Erickson:2000af}. In the planar limit $N_{c}\to \infty$, with fixed 't Hooft coupling 
$\widehat\lambda = g^{2}_{\rm YM}\, N_{c}$~\footnote{
We introduce a slightly unconventional hat in the notation for the 't Hooft coupling
to avoid confusion with the coupling $\lambda$ in (\ref{1.1}) that we prefer to keep in order to match 
the conventions in the mathematical physics literature.}, we can obtain 
the quark-antiquark potential $-\frac{1}{r}\,\Omega(\widehat\lambda)$  from 
the expectation value of the associated Wilson loop. The function $\Omega(\widehat\lambda)$ is known 
at 3 loops at weak coupling
\cite{Erickson:1999qv,Pineda:2007kz,Correa:2012nk,Bykov:2012sc,Stahlhofen:2012zx,Prausa:2013qva,Drukker:2011za}  and at one loop at strong coupling 
\cite{Maldacena:1998im,Rey:1998ik,Forini:2010ek,Chu:2009qt}. 
This construction can be extended by introducing an angular 
parameter $\vartheta$ that takes into account the relative flavours of the quark and antiquark.
The generalized potential $\Omega(\widehat \lambda, \vartheta)$ has been recently studied in \cite{Gromov:2016rrp}
by means of the Quantum Spectral Curve methods developed in  \cite{Gromov:2015dfa}. One of the remarkable
results of  \cite{Gromov:2016rrp} is the  analytic weak coupling expansion of $\Omega(\widehat\lambda, \vartheta)$
up to 7 loops, {\em i.e.} at order $\mc O(\lambda^{7})$.
A particularly interesting limit is the double scaling regime
$\vartheta\to i\,\infty$ with fixed $\widehat \lambda\,e^{-i\,\vartheta}$. This limit resums the ladder diagrams in the gauge theory and the quark-antiquark potential is obtained as the ground state energy of a  1d Schr\"odinger equation with 
potential in the class (\ref{1.4}). In the notation of (\ref{1.1}), the 
problem addressed in \cite{Gromov:2016rrp} corresponds to the specific potential 
\be
\la{1.6}
V(x) \equiv V_{\rm I}(x) = -\frac{1}{1+x^{2}},
\ee
with Schr\"odinger coupling $\lambda = \frac{\widehat{\lambda}\,e^{-i\,\vartheta}}{16\pi^{2}}$
and identification $E_{0} = -\frac{1}{4}\,\Omega^{2}(\lambda)$.
The first three terms of the expansion of $\sqrt{-E_{0}}$ may be written \cite{Gromov:2016rrp} 
\be
\label{1.7}
\begin{split}
\sqrt{-E_{0}(\lambda)} &= \frac{\pi}{2}\,\lambda+\pi\,(\log\lambda+2\,\mc L-1)\,\lambda^{2}\\
&+\pi\,\bigg[\log^{2}\lambda+(4\mc L+1)\,\log\lambda-\frac{1}{12}(-48\,\mc L^{2}-24\,\mc L+\pi^{2}+42)\bigg]\,\lambda^{3}+\cdots,
\end{split}
\ee
where $\mc L=\log\sqrt{2\,e^{\gamma_{\rm E}}\,\pi}$, and $\gamma_{\rm E}$ denotes the Euler-Mascheroni
constant. Inspection of the $\mc O(\lambda^{7})$ 
additional terms in  (\ref{1.7}) shows that each power $\lambda^{n}$ comes
together with all logarithms $\log^{m}\lambda$ with $0\le m\le n-1$.
Also, the leading and next-to-leading logarithms seem to be captured by the resummation  Ansatz
\be
\la{1.8}
\left. \sqrt{-E_{0}(\lambda)}\right|_{\rm NLO} = \frac{\pi}{2}\,\lambda\,
e^{2\,\lambda\,\text{L}}\,(1+6\,\lambda^{2}\,\text{L}), \qquad \text{L}=\log\lambda+2\mc L-1,
\ee
whose leading part is a simple exponentiation.
The expansion (\ref{1.7}) and the resummation (\ref{1.8}) pose several interesting questions. The first is whether 
the structure of (\ref{1.7}) 
is general, {\em i.e.} if it is true that -- at least at third order -- for all potentials in the class (\ref{1.4})
one has 
\be
\label{1.9}
\sqrt{-E_{0}(\lambda)} = c_{1,0}\,\lambda+(c_{2,1}\,\log\lambda+c_{2,0})\,\lambda^{2} + 
(c_{3,2}\,\log^{2}\lambda+c_{3,1}\,\log\lambda+c_{3,0})\,\lambda^{3} + \cdots.
\ee
The second question concerns the general validity of resummation formulas like (\ref{1.8}). 
In the context of the static quark-antiquark potential, resummation of leading and next-to-leading logarithms
has been performed by a Renormalization Group (RG) analysis in 
\cite{Pineda:2007kz,Stahlhofen:2012zx}. In particular, leading logarithms simply exponentiate 
and, since they are computed  exactly by the ladder approximation, this explain the first term in  (\ref{1.8}). 
Nevertheless, it would be interesting to understand 
whether such  exponentiation is expected for a general potential and can be determined at the level of the 
Schr\"odinger equation.
This question is particularly intriguing because in gauge theories the RG exponentiation -- at leading order --
encodes the factorization scale dependence that remains after cancellation between real and virtual infrared
divergences. In the Schr\"odinger equation, this deep machinery is somewhat absent and 
exponentiation of logarithms of $\lambda$ is less clear and more elusive.
%

\medskip
\noindent
The plan and content of this paper is the following. In Sec.~(\ref{sec:solvable}), 
we discuss a fully solvable potential in the class (\ref{1.4}) where we provide the expansion of the
binding energy at high order showing that it admits the structure in (\ref{1.9}) as well as an exponentiation
formula similar to (\ref{1.8}). In Sec.~(\ref{sec:second-order}), we present a first simple extension
of the result (\ref{1.5}), by computing the subleading coefficient $c_{2,0}$ in (\ref{1.9}). The derivation 
is rather straightforward, but cannot be easily extended at higher orders. 
To overcome these difficulties, we apply in Sec.~(\ref{sec:MAE}) the general approach of matched asymptotic expansions
to our  problem. As a preparation,  Sec.~(\ref{subsec:toy}) is devoted to discuss
a toy-model that captures the essence of the method. Then, in Sec.~(\ref{subsec:real}), 
we apply it to the determination of (\ref{1.9}) and we compute the complete third order expansion 
for a general potential. Our main result, summarized in (\ref{4.39}) and  (\ref{4.40}), is 
 tested on various solvable and non-solvable examples in Sec.~(\ref{sec:checks}). In particular, 
Sec.~(\ref{subsec:numerical}) presents several high precision numerical tests of the third order expansions to appreciate
its accuracy at small coupling, despite being only  asymptotic. As a final test, we discuss in Sec.~(\ref{sec:susy})
a consistency check of our expansion, when applied to a supersymmetric potential
arising in the study of the quark-antiquark potential in $\mc N=6$ ABJ(M) theory. Finally, 
our third order expansion suggests  exponentiation of the leading logarithms in the form
(see the first term of (\ref{1.8}) as a special case)
\be
\la{1.10}
\sqrt{-E_{0}(\lambda)} = -\frac{1}{2}\,\lambda\,e^{2\,a\,\lambda\log\lambda}\,\int_{\mathbb R}dx\, V(x) + \text{subleading logarithms}.
\ee
In Sec.~(\ref{sec:resummation}), we prove
that this result holds at all orders. Finally, Sec.~(\ref{sec:conclusions}) summarises our conclusions and 
outlines possible 
further directions. 

\section{A solvable example}
\label{sec:solvable}

The potential (\ref{1.6}) may be treated by the Quantum Spectral Curve as a special limit.
To improve our knowledge, we also analyze the following solvable case discussed in  
\cite{Blankenbecler:1977pf,van1978bound}
\be
\label{2.1}
V(x) \equiv V_{\rm II}(x) = -\frac{1}{4}\,\frac{1}{(1+|x|)^{2}}.
\ee
Its ground state energy $\alpha\equiv \sqrt{-E_{0}(\lambda)}$ obeys the exact equation 
\be
\label{2.2}
\alpha\,\frac{\partial}{\partial \alpha}\,K_{\nu}(\alpha)+ \frac{1}{2}\,K_{\nu}(\alpha) = 0, \qquad \nu = \frac{1}{2}\,\sqrt{1-\lambda},
\ee
where $K_{\nu}(\alpha)$ are modified Bessel functions of the second kind.
Expanding (\ref{2.2}) at small $\lambda$, we get the asymptotic series
\be
\label{2.3}
\begin{split}
\sqrt{-E_{0}(\lambda)} &=\frac{1}{4}\,\lambda+\frac{\text{L}}{8}\,\lambda^{2}
+\bigg(
\frac{\text{L}^2}{32}+\frac{5   \text{L}}{32}-\frac{11}{128}\bigg)\,\lambda^{3}\\
&+\bigg(
\frac{\text{L}^3}{192}+\frac{7 \text{L}^2}{64}+\frac{21
   \text{L}}{256}+\frac{1}{768} (14 \,\zeta_{3}-71)\bigg)\,\lambda^{4}\\
   &+\bigg(\frac{\text{L}^4}{1536}+\frac{35 \text{L}^3}{768}+\frac{175
   \text{L}^2}{1024}+\frac{\text{L} (28 \,\zeta_{3}-89)}{3072}+\frac{2352 \,\zeta_{3}-5833}{73728}\bigg)\,\lambda^{5}\\
   &+\bigg(\frac{\text{L}^5}{15360}+\frac{23
   \text{L}^4}{1536}+\frac{859 \text{L}^3}{6144}+\frac{\text{L}^2 (7 \,\zeta_{3}+444)}{3072}\\
   &+\frac{\text{L} (5376 \,\zeta_{3}-19105)}{147456}+\frac{5355 \,\zeta_{3}+558 \,\zeta_{5}-9176}{184320}\bigg)\,\lambda^{6}+\cdots,
      \end{split}
\ee
where $\text{L} = \log\lambda-\log 2 + \gamma_{\rm E}+\tfrac{1}{2}$. All the leading 
$\lambda^{n+1}\,\text{L}^{n}$ and subleading $\lambda^{n+2}\,\text{L}^{n}$ logarithms
are captured by the following formula (see App.~(\ref{App:resum}) for a proof)
\be
\label{2.4}
\left. \sqrt{-E_{0}(\lambda)}\right|_{\rm NLO} = \tfrac{1}{4}\,\lambda\,\bigg[
e^{\frac{1}{2}\,\lambda\,\text{L}}\,\bigg(
1-\tfrac{1}{4}\,\lambda+\tfrac{3}{8}\,\lambda^{2}\,\text{L}
\bigg)+\tfrac{1}{4}\,\lambda\,e^{\frac{3}{2}\,\lambda\,\text{L}}\bigg].
\ee
The first term in the square bracket  is similar to the resummation (\ref{1.8}). 
Some differences in the other terms are not surprising  because 
here the $\mc O(x^{-3})$ subleading term at infinity is non zero, and may be important.
Collecting data from  (\ref{1.7}) and (\ref{2.3}), we obtain the  reference table Tab.~(\ref{tab1})
for the coefficients appearing in  (\ref{1.9}) up to third order:
\begin{table}[H]
\be
\notag
\renewcommand{\arraystretch}{1.4}
\begin{array}{|c||c|cc|ccc|}
\hline
& c_{1,0} & c_{2,1} & c_{2,0} & c_{3,2} & c_{3,1} & c_{3,0} \\
\hline
V_{\rm I}(x) & \frac{\pi}{2} & \pi & \pi\,(2\,\mc L-1) & \pi & \pi\,(4\,\mc L+1) & 
-\frac{1}{12}(-48\,\mc L^{2}-24\,\mc L+\pi^{2}+42)\\
\hline
V_{\rm II}(x) & \frac{1}{4} & \frac{1}{8} & \frac{1}{8}\,(\gamma_{\rm E}-\log 2+\tfrac{1}{2})
& \tfrac{1}{32} & \tfrac{1}{16}(3+\gamma_{\rm E}-\log 2) & \tfrac{1}{32}(\gamma_{\rm E}-\log 2)(\gamma_{\rm E}-\log 2+6) \\
\hline
\end{array}
\ee
\caption{Summary table showing the coefficient of the asymptotic expansion of the binding energy 
at third order for the test case potentials $V_{\rm I}(x)$ and $V_{\rm II}(x)$.}
\label{tab1}
\end{table}
This is what we want to be able to compute 
for a general potential  in the class (\ref{1.4}). Of course, the (leading) second order rigorous result (\ref{1.5}) is 
fully consistent with the entries in Tab.~(\ref{tab1}).

\section{The complete second order expansion}
\label{sec:second-order}

A first simple extension of the result (\ref{1.5}) obtained in   \cite{Blankenbecler:1977pf}  concerns the 
determination of the coefficient $c_{2,0}$ in (\ref{1.9}).
The analysis in  \cite{Blankenbecler:1977pf} 
is based on the results of \cite{MR0404846} that imply that 
the ground state energy $E_{0}=-\alpha^{2}$ is obtained from the condition 
\be
\label{3.1}
\det(1+\lambda\,K_{\alpha}) = 0,
\ee
where $K_{\alpha}$ is the Birman-Schwinger operator \cite{reed1978analysis}
\be
\label{3.2}
K_{\alpha}(x,y) = \frac{1}{2\,\alpha}|V(x)|^{1/2}\,e^{-\alpha\,|x-y|}\,|V(y)|^{1/2}\,\text{sign}(V(y)).
\ee
We can expand (\ref{3.1}) as follows
\be
\label{3.3}
\begin{split}
\det(1+\lambda K_{\alpha}) &= e^{\text{tr}\log(1+\lambda K_{\alpha})} =  1+ \lambda\,\text{tr}K_{\alpha}
+\frac{\lambda^{2}}{2}\,\bigg[(\text{tr}K_{\alpha})^{2}-\text{tr}K_{\alpha}^{2}\bigg]\\
&+\frac{\lambda^{3}}{6}\bigg[
(\text{tr}K_{\alpha})^{3}-3\,\text{tr}K_{\alpha}\,\text{tr}K_{\alpha}^{2}+2\,\text{tr}K_{\alpha}^{3}
\bigg]+\cdots.
\end{split}
\ee
At second order,  (\ref{3.3}) may be written in the simple form  \footnote{
An alternative expression for the second order term is 
\be\notag
\int_{-\infty}^{\infty} dx\, dy\,V(x) \,V(y)\,(1-e^{-2\,\alpha\,|x-y|}) = 
2\,\int_{-\infty}^{\infty} dx\,\int_{0}^{\infty}dz\, V(x) \,V(x+z)\,(1-e^{-2\,\alpha\,z}).
\ee
}
\be
\label{3.4}
0 = 1+\frac{\lambda}{2\,\alpha}\,\int_{\mathbb R}dx V(x) +\frac{\lambda^{2}}{8\,\alpha^{2}}
\int_{\mathbb R} dx\, \int_{\mathbb R} dy\,V(x) \,V(y)\,(1-e^{-2\,\alpha\,|x-y|})+\cdots.
\ee
The rigorous analysis of \cite{Blankenbecler:1977pf}  leads to the result (\ref{1.5}) for the $\lambda$
and $\lambda^{2}\,\log\lambda$ term. If we want to capture  
the $\lambda^{2}$ term  (without logarithmic enhancement),
we need to study in details the small $\alpha$ limit of the second order correction
\be
\label{3.5}
F(\alpha) = \int_{\mathbb R} dx\,\int_{\mathbb{R}} dy\,V(x) \,V(y)\,(1-e^{-2\,\alpha\,|x-y|}).
\ee
To this aim, we define the Fourier transform of $V(x)$ with the convention 
\be
\label{3.6}
\widetilde V(\omega) = \frac{1}{(2\pi)^{1/2}}\int_{\mathbb R} dx\, V(x)\,e^{i\,\omega\,x},
\ee
and easily obtain  the alternative form of $F(\alpha)$
\be
\label{3.7}
\begin{split}
F(\alpha) &= 2\,\pi\,\int_{\mathbb R}d\omega\,|\widetilde V(\omega)|^{2}
\bigg[\delta(\omega)-\frac{2\,\alpha}{\pi\,(4\,\alpha^{2}+\omega^{2})}\bigg] 
= 2\,\int_{\mathbb R}d\omega\,\frac{|\widetilde V(0)|^{2}-|\widetilde V(2\,\alpha\,\omega)|^{2}}{1+\omega^{2}}
\end{split}.
\ee
For a specific potential, one can study this integral for $\alpha\to 0$
 \footnote{For instance, if the Mellin transform of 
$|\widetilde V(\omega)|^{2}$ is known, then it is easy to obtain the small $\alpha$ asymptotic expansion of $F(\alpha)$.},
but here we want to discuss (\ref{3.7}) in  general terms. One straightforward approach is the following.
Since $V(x)$ is real, $\widetilde V(\omega)^{*} = \widetilde V(-\omega)$ and 
$U(\omega) \equiv |\widetilde V(\omega)|^{2}$
 is even. The expression (\ref{3.7}) can be split as follows
\be
\label{3.8}
\begin{split}
F(\alpha) &= 4\,\int_{0}^{\infty}d\omega\,\frac{U(0)-U(2\,\alpha\,\omega)}{1+\omega^{2}}
= 4\,\bigg(\int_{0}^{1/(2\alpha)}+\int_{1/(2\alpha)}^{\infty}\bigg)\,d\omega\,\frac{U(0)-U(2\,\alpha\,\omega)}{1+\omega^{2}} \\
&= 8\,\alpha\,\bigg(\int_{0}^{1}+\int_{1}^{\infty}\bigg)\,d\omega\,\frac{U(0)-U(\omega)}{4\,\alpha^{2}+\omega^{2}}
\end{split}
\ee
The first integral is computed by adding and subtracting in the numerator $\omega\,U'(0^{+})$ and splitting 
the elementary part that can be computed explicitly. In what remains, as well as in the second integral, 
we can safely send $\alpha\to 0$ if we are interested in the $\mc O(\alpha\log\alpha)$ and $\mc O(\alpha)$ 
terms of $F(\alpha)$.
In conclusion,
\be
\label{3.9}
\begin{split}
F(\alpha) &= 8\,\alpha\,\bigg[
U'(0^{+})\,(\log\alpha+\log 2)\\
&\qquad +\int_{0}^{1}\frac{U(0)+U'(0^{+})\,\omega-U(\omega)}{\omega^{2}}
+\int_{1}^{\infty}\frac{U(0)-U(\omega)}{\omega^{2}}
\bigg]+\mc O(\alpha^{2}).
\end{split}
\ee
Given
\be
\label{3.10}
F(\alpha) = \alpha\,(F_{1}\,\log\alpha + F_{0})+\mc O(\alpha^{2}),
\ee
we obtain from (\ref{3.4})
\be
\label{3.11}
0 = 1+\frac{\lambda}{2\,\alpha}\int_{\mathbb R}dx\, V+\frac{\lambda^{2}}{8\,\alpha}(F_{1}\,\log\alpha+F_{0}) +\cdots.
\ee
Denoting for brevity
\be
\label{3.12}
\boxed{
\overline V = \int_{\mathbb R}dx\, V(x), 
}
\ee
and inverting (\ref{3.11}), we find
\be
\label{3.13}
\alpha = -\frac{\overline V}{2}\,\lambda-\frac{\lambda^{2}}{8}\,
\bigg[F_{1}\,\log\lambda+ F_{0}+F_{1}\,\log\left(-\tfrac{1}{2}
\,\overline V\right)\bigg]+\cdots.
\ee
In conclusion, we find the second order terms in the expansion (\ref{1.9}) with the following coefficients 
\be
\label{3.14}
\boxed{
\begin{split}
c_{1,0} &= -\tfrac{1}{2}\,\overline V, \\
c_{2,1} &= -U'(0^{+}), \\
c_{2,2} &= -U'(0^{+})\,\log(-\overline V)-\int_{0}^{1}\frac{U(0)+U'(0^{+})\,\omega-U(\omega)}{\omega^{2}}
-\int_{1}^{\infty}\frac{U(0)-U(\omega)}{\omega^{2}}.
\end{split}
}
\ee
Notice that, to match (\ref{3.3}), we need  to show that $U'(0^{+}) = a\,\overline V$.
This is slighlty non trivial. For an even potential
$\widetilde V(\omega)$ is also even, and then $U'(0^{+}) = 2\,\widetilde V(0)\,\widetilde V'(0^{+})$.
Then,
\be
\label{3.15}
\begin{split}
\widetilde V'(\omega) &= \frac{i}{\sqrt{2\pi}}\int_{\mathbb R} x\,V(x)\,e^{i\,\omega\,x} 
= 
\frac{i}{\sqrt{2\pi}}\int_{\mathbb R} x\,\bigg[V(x)+\frac{a}{1+x^{2}}\bigg]\,e^{i\,\omega\,x} 
-\frac{i\,a}{\sqrt{2\pi}}\int_{\mathbb R} \frac{x}{1+x^{2}}\,e^{i\,\omega\,x}.
\end{split}
\ee
In the first integral we can safely send $\omega\to 0$ and then it vanishes because it is odd under $x\to -x$. The remaining piece gives
\be
\label{3.16}
\begin{split}
\widetilde V'(0^{+}) &= 
\frac{a}{\sqrt{2\pi}}\lim_{\omega\to 0^{+}}\int_{\mathbb R} \frac{x\,\sin(\omega\,x)}{1+x^{2}} = 
\lim_{\omega\to 0^{+}}\frac{a}{\sqrt{2\pi}}\frac{\pi\,\omega\,e^{-|\omega|}}{|\omega|} = a\,\sqrt\frac{\pi}{2}.
\end{split}
\ee
Finally, using $\widetilde V(0) = \frac{1}{\sqrt{2\pi}}\,\overline V$, we match (\ref{3.3}).

\subsection{Check with the potential $V_{\rm I}(x) = -\frac{1}{1+x^{2}}$}

In this case, 
\be
\label{3.17}
\overline V = -\pi, \qquad U(\omega) = \frac{\pi}{2}\,e^{-2\,\omega}, \qquad U'(0^{+}) = -\pi.
\ee
Also, we can compute \footnote{Here, the exponential integral function is 
\be\notag
\text{Ei}(z) = -\int_{-z}^{\infty}\frac{e^{-t}}{t}\,dt.
\ee
}
\be
\label{3.18}
\begin{split}
\int_{0}^{1}d\omega\,\frac{U(0)+U'(0^{+})\,\omega-U(\omega)}{\omega^{2}} &= \frac{1}{2} 
\pi  \left(2\, \text{Ei}(-2)+1+\frac{1}{e^2}-2
   \gamma_{\rm E} -2\,\log 2\right), \\
\int_{1}^{\infty}d\omega\,\frac{U(0)-U(\omega)}{\omega^{2}} &=  \frac{1}{2} \pi  \left(-2
   \text{Ei}(-2)+1-\frac{1}{e^2}\right).
\end{split}
\ee
Replacing in (\ref{3.14}), we obtain 
\be
\label{3.19}
\begin{split}
c_{1,0} &= \tfrac{\pi}{2}, \qquad
c_{2,1} = \pi, \qquad
c_{2,2} = \pi\,(-1+\gamma_{E}+\log 2+\log \pi) = \pi\,(2\,\mc L-1),
\end{split}
\ee
in full agreement with the entries in Tab.~(\ref{tab1}).

\subsection{Check with the potential $V_{\rm II}(x) = -\frac{1}{4\,(1+|x|)^{2}}$}

In this case, 
\be
\label{3.20}
\widetilde V(\omega) = -\frac{2-\omega  (2 \text{Ci}(\omega ) \sin \omega +(\pi -2
   \text{Si}(\omega )) \cos \omega )}{4 \sqrt{2 \pi }}
\ee
and using $\widetilde V(\omega) = -\frac{1}{2 \sqrt{2 \pi }}+\frac{1}{4} \sqrt{\frac{\pi }{2}}
   \omega +\cdots$, we find 
\be
\label{3.21}
\overline V = -\frac{1}{2}, \qquad U(0) = \frac{1}{8\,\pi}, \qquad U'(0^{+}) = -\frac{1}{8}.
\ee
Also, we can compute the relevant non trivial integrals numerically at high precision finding
\be
\label{3.22}
\begin{split}
I_{1} &= \int_{0}^{1}d\omega\,\frac{U(0)+U'(0^{+})\,\omega-U(\omega)}{\omega^{2}}   = 
-0.172399851825784556337918255\dots,\\
I_{2} &= \int_{1}^{\infty}d\omega\,\frac{U(0)-U(\omega)}{\omega^{2}} =
0.03774789371309294876210\dots,
\end{split}
\ee
Remarkably, we checked that up to 100 digits we have 
\be
\label{3.23}
I_{1}+I_{2} = -\tfrac{1}{16}(1+2\,\gamma_{\rm E}).
\ee
Replacing in (\ref{3.14}), we obtain 
\be
\label{3.24}
\begin{split}
c_{1,0} &= -\tfrac{1}{4}, \qquad
c_{2,1} = -\tfrac{1}{8}, \qquad
c_{2,2} = \tfrac{1}{16}(1+2\,\gamma_{\rm E}-2\log 2).
\end{split}
\ee
again in full agreement with  Tab.~(\ref{tab1}).

\subsection{Leading logarithms at third order}

One may attempt to apply this approach to 
work  out the third order contribution to $\sqrt{-E_{0}}$, but this  is definitely a non-trivial task,
although the leading order term $\sim \lambda^{3}\,\log^{2}\lambda$ is heuristically derived
in App.~(\ref{App:3naive}). In the next section, we shall introduce an alternative method 
that is powerful enough to fully compute the third order expansion and, in principle, also higher order 
terms.

\section{Matched Asymptotic Expansion, the complete third order}
\label{sec:MAE}

A completely different approach is based on the method of matched asymptotic expansions for 
boundary layer problems \cite{hinch1991perturbation,miller2006applied}.
Although the idea is the same as in the standard examples of singular perturbation of 
differential equations \cite{chen1996renormalization}, 
the specific application to the present setup is rather different and somewhat unusual. \footnote{
The application of matched asymptotic expansion methods to the Schr\"odinger equation in the semiclassical
limit is of course a text-book topic, see for instance  \cite{singh1983singular} for a clean 
pedagogical presentation. In that context, matching is applied around the inversion points.}
Earlier studies in the context of the quark-antiquark potential may be found 
in  \cite{Klebanov:2006jj,Brower:2006hf}. For the potential  $V_{\rm I}(x)$, this method
has been successfully exploited in \cite{Correa:2012nk}. The method is fully general and we shall 
present its application for a generic potential in the class (\ref{1.4}). A recent application 
to a different kind of Schr\"odinger problem may be found in \cite{rosales2015asymptotic}. \footnote{
The main new complication to be discussed here is the presence of non-analytic terms in a boundary problem
eigenvalue. This is a feature that, for instance, is absent in the analysis of the Mathieu equation in \cite{chen1996renormalization}.}
We now first present a simple example illustrating the procedure and then apply the method to our case.

\subsection{A toy model}
\label{subsec:toy}

\begin{figure}[H]
\begin{center}
\hskip 1cm \includegraphics[scale=0.8]{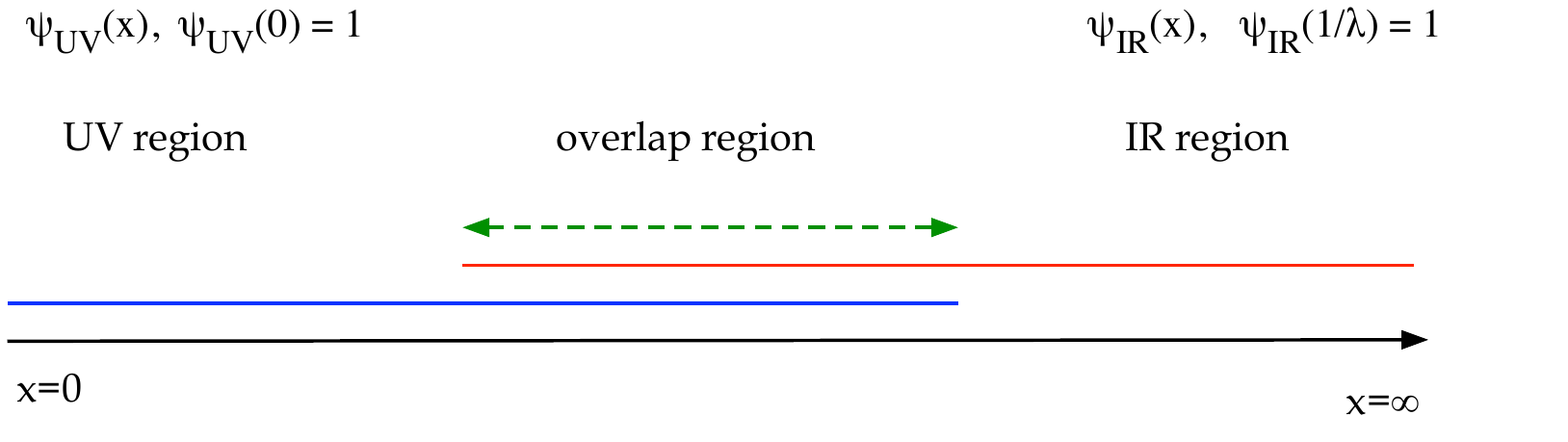}
\caption{Qualitative picture illustrating the Matched Asymptotic Expansion procedure.}
\label{fig:MAE}
\end{center}
\end{figure}

Let us consider the first order boundary problem
\be
\la{4.1}
\begin{split}
\psi'(x) &= \bigg(\alpha+\frac{\lambda}{1+x}\bigg)\,\psi(x), \\
\psi(0) &= 1, \qquad \psi(\lambda^{-1}) = 1.
\end{split}
\ee
Although apparently contrived, this problem captures the essential features of our problem. We are 
interested in the $\lambda\to 0^{+}$ limit and there are two boundary 
conditions, one in the UV region, at small $x$, and one in the IR region, at large $x$. The two boundary conditions
determine $\alpha$ as a function of $\lambda$. Indeed, the exact solution of (\ref{4.1}) before imposing the boundary condition
at $x=\lambda^{-1}$ is 
\be
\la{4.2}
\psi(x) = e^{\alpha\,x}\,(1+x)^{\lambda}.
\ee
Imposing $\psi(\lambda^{-1})=1$, we obtain  the following asymptotic expansion \footnote{
We remark that in this simple example, 
the non-analyticity due to logarithms is limited to the first term, while the remainder is actually a power series
with unit convergence radius.}
\be
\la{4.3}
\alpha = -\lambda^{2}\,\log\bigg(1+\frac{1}{\lambda}\bigg) =\lambda^{2}\,\log\lambda-\lambda^{3}+\frac{1}{2}\lambda^{4}+\cdots,
\ee
that is the result we want to derive by an alternative route. To this aim, we apply the ideas of matched asymptotic expansions and look  for two approximate solutions of (\ref{4.1}). The first is $\psi_{\rm IR}(x)$ and is valid 
around $x=\infty$, with $\psi_{\rm IR}(\lambda^{-1})=1$. The second is $\psi_{\rm UV}(x)$ and must be used
around $x=0$, with $\psi_{\rm UV}(0)=1$. The two solutions must be compared by imposing the symbolic 
matching
\be
\la{4.4}
\lim_{x\to \infty} \psi_{\rm UV}(x) = \lim_{x\to 0} \psi_{\rm IR}(x).
\ee
For this procedure to make sense, the precise meaning of the relation (\ref{4.4}) is that we need 
$\psi_{\rm UV}(x) \simeq \psi_{\rm IR}(x)$ in an intermediate overlap region, as shown in 
Fig.~(\ref{fig:MAE}). Besides, in our 
application the matching will be imposed perturbatively in $\lambda$. 

Following this approach, the first step is to consider the approximate equation 
that is obtained at large $x$ and solve it with the IR boundary condition: 
\be
\la{4.5}
\begin{split}
\psi'_{\rm IR}(x) &= \bigg[\alpha+\lambda\,\bigg(\frac{1}{x}-\frac{1}{x^{2}}+\frac{\xi_{3}}{x^{3}}
-\frac{\xi_{4}}{x^{4}}\bigg)\bigg]\,\psi_{\rm IR}(x), \\
\psi_{\rm IR}(\lambda^{-1}) &= 1.
\end{split}
\ee
We have kept four terms in the large $x$ expansion of the potential. This is instructive with the aim of keeping
track of  the role of the various corrections. In other words, the coefficients $\xi_{3}, \xi_{4}$ are useful to identify 
terms that depend on including ($\xi=1$) or excluding ($\xi=0$) a certain subleading part of $V$.
The solution of (\ref{4.5}) is 
\be
\la{4.6}
\psi_{\rm IR}(x) = \lambda^{\lambda}\,x^{\lambda}\,\exp\bigg[
-\frac{\alpha}{\lambda}+\alpha\,x+\lambda\,\bigg(\frac{1}{x}-\frac{\xi_{3}}{2x^{2}}+\frac{\xi_{4}}{3x^{3}}\bigg)
-\lambda^{2}+\frac{\lambda^{3}}{2}\,\xi_{3}-\frac{\lambda^{4}}{3}\,\xi_{4}\bigg].
\ee
Now, we make the educated guess \footnote{
The appearance of a double expansion in $\lambda$ and $\log\lambda$ is familiar in the context of 
matched asymptotic techniques where logarithmic terms are commonly named as {\em switchback} corrections
 \cite{holzer2014analysis}.}
\be
\la{4.7}
\alpha = A\,\lambda^{2}\,\log\lambda+B\,\lambda^{3}+\cdots,
\ee
and expand (\ref{4.6}) at small $\lambda$, and fixed $x$. This is an expansion in powers of $\lambda$ with possible 
powers of $\log\lambda$. Explicitly, we find
\be
\la{4.8}
\psi_{\rm IR}(x) = 1+\lambda\,\psi_{\text{IR}, 1}(x)+\lambda^{2}\,\psi_{\text{IR}, 2}(x)+\cdots,
\ee
with 
\be
\la{4.9}
\begin{split}
\psi_{\rm IR, 1}(x) &= (1-A)\, \log \lambda +\log x +\frac{1}{x}-\frac{\xi _3}{2 x^2}+\frac{\xi _4}{3
   x^3}, \\
\psi_{\rm IR, 2}(x) &= A\, x\, \log \lambda +\frac{1}{2} [(A-1)^2\, \log ^2\lambda-2
 \,  (A-1)\, \log \lambda  \log x-2\, (B+1)+\log^2 x]\\
   &+\frac{\log x+(1-A)\,\log \lambda}{x}+\frac{1-\xi _3\, [\log x +(1-A)\, \log \lambda]}{2 x^2}\\
   &+\frac{2\, \xi _4\, [\log x + (1-A)\, \log \lambda]-3\, \xi _3}{6 x^3}+\frac{\frac{\xi _3^2}{8}+\frac{\xi
   _4}{3}}{x^4}-\frac{\xi _3 \xi _4}{6 x^5}+\frac{\xi _4^2}{18
   x^6}.
\end{split}
\ee
In the UV region, we consider the unmodified initial problem, but impose only the boundary condition at $x=0$
\be
\la{4.10}
\begin{split}
\psi_{\rm UV}'(x) &= \bigg(\alpha+\frac{\lambda\,x}{1+x}\bigg)\,\psi_{\rm UV}(x), \\
\psi_{\rm UV}(0) &= 1.
\end{split}
\ee
To match (\ref{4.8}), we solve (\ref{4.10}) pertubatively in $\lambda$, taking into account (\ref{4.7}),
\be
\la{4.11}
\psi_{\rm UV}(x) = 1+\lambda\,\psi_{\rm UV, 1}(x)+\lambda^{2}\,\psi_{\rm UV, 2}(x) + \cdots, \qquad 
\psi_{\text{UV}, n}(0)=0.
\ee
This gives immediately
\be
\la{4.12}
\begin{split}
\psi_{\rm UV, 1}(x) &=\log(x+1), \\
\psi_{\rm UV, 2}(x) &=A\,x\,\log \lambda+\frac{1}{2}\,\log^{2}(x+1).
\end{split}
\ee
Finally, we attempt to match the two solutions (\ref{4.8}) and (\ref{4.11}) by evaluating the ratio 
\be
\la{4.13}
R(x, \lambda) = \frac{\psi_{\rm IR}(x)}{\psi_{\rm UV}(x)} = 1+\lambda\, R_{1}(x, \lambda)
+\lambda^{2}\, R_{2}(x, \lambda)+\cdots.
\ee
The contribution from the UV
region must be expanded at large $x$ where the overlap region is located. Doing so, we  obtain 
at first order
\be
\la{4.14}
R_{1}(x, \lambda) = (1-A)\,\log \lambda+\frac{1-\xi_{3}}{2\,x^{2}}
+\frac{\xi _4-1}{3 x^3}+\frac{1}{4 x^4}-\frac{1}{5
   x^5}+\cdots.
\ee
Thus, we see that matching ($R_{n}=0$) requires $A=1$ in agreement with (\ref{4.3}). This does not depend on the
$\xi$-terms, {\em i.e.} the $x^{-3}$ and $x^{-4}$ corrections to $V$ included in (\ref{4.5}). The correct choice 
$\xi_{3}=\xi_{4}=1$ cancels the various corrections in (\ref{4.14}), but we repeat that this is irrelevant to 
fix $A$. With the choice $A=1$, we can inspect $R_{2}(x, \lambda)$ and find
\be
\la{4.15}
R_{2}(x, \lambda) = -B-1+\frac{\left(\xi _3-1\right){}^2}{8 x^4}-\frac{\left(\xi
   _3-1\right) \left(\xi _4-1\right)}{6 x^5}+\frac{4 \xi _4^2-8
   \xi _4-9 \xi _3+13}{72 x^6}+\frac{6 \xi _3+5 \xi _4-11}{60
   x^7}+\cdots.
\ee
Again, the coefficient $B$ is fixed at $B=-1$ in agreement with (\ref{4.3}). The first two subleading corrections in (\ref{4.15}) vanish if we include the $x^{-3}$ correction in (\ref{4.5}). Doing so, {\em i.e.} taking $\xi_{3}=1$, 
we get
\be
\la{4.16}
\left. R_{2}(x, \lambda)\right|_{B=-1, \ \xi_{3}=1} = \frac{\left(\xi _4-1\right){}^2}{18 x^6}+\frac{\xi _4-1}{12
   x^7}+\frac{47-32 \xi _4}{480 x^8}+\frac{10 \xi _4-19}{180
   x^9}+\cdots,
\ee
showing that the inclusion of the $x^{-4}$ correction in (\ref{4.5}), {\em i.e.} taking $\xi_{4}=1$,  cancel two additional subleading terms.

\subsection{Application to the third order correction}
\label{subsec:real}

We now apply to our main problem
the procedure that worked for the toy model  (\ref{4.1}).

\subsubsection{Analysis of the infrared region}

Let us consider an even potential admitting the expansion for $x\to +\infty$
\be
\label{4.17}
V(x) = -\frac{a}{x^{2}}-\frac{b}{x^{3}}+\mc O(x^{-4}).
\ee
The large $x$ equation
\be
\label{4.18}
-\psi''(x)+\bigg(\alpha^{2}-\frac{a\,\lambda}{x^{2}}\bigg)\,\psi(x)=0,
\ee
has the following solution decaying exponentially at infinity
\be
\label{4.19}
\psi(x) = \psi^{(1)}(x) =  \sqrt\frac{2\,\alpha\,x}{\pi}\,\text{K}_{\frac{1}{2}\,\sqrt{1-4\,a\,\lambda}}(\alpha\,x).
\ee
The second independent solution is 
\be
\label{4.20}
\psi^{(2)}(x) =  \sqrt\frac{2\,\alpha\,x}{\pi}\,\text{I}_{\frac{1}{2}\,\sqrt{1-4\,a\,\lambda}}(\alpha\,x).
\ee
Taking into account the next to leading correction to the potential at large $x$, the IR solution is
\be
\label{4.21}
\psi_{\rm IR}(x) = \psi^{(1)}(x) +\frac{\pi\,\lambda}{2\,\alpha}\,\bigg[
b\, \psi^{(1)}(x)\,\int^{x} dx' \,\frac{\psi^{(2)}(x')\,\psi^{(1)}(x')}{(x')^{3}}-b\,\psi^{(2)}(x)\,\int^{x}dx \,\frac{
\psi^{(1)}(x')^{2}}{(x')^{3}}
\bigg]+\dots.
\ee
As we explained, we want to expand this solution in the limit $\alpha\,x\to 0$, and then $\lambda\to 0$. It is convenient to set 
$\alpha = A(\lambda)\, \lambda$ and we find ($c_{n}$ are constant terms dependent of $A, a, b, \lambda$, but independent of $x$)
\be
\label{4.22}
\begin{split}
\psi_{\rm IR}(x) &= 1+\lambda\,\bigg[-\frac{b}{2 x}+a \log x-A x+c_{1}\bigg]+\\
&+\lambda^{2}\,\bigg[\frac{A^2 x^2}{2}+
x \bigg(\frac{A \left(2 a^2 \text{L}_{A}-4 a^2-A b\right)}{2 a}+a A
   \log x\bigg)\\
   &+\frac{1}{2} \log x \left(2 a^2 \text{L}_{A}+2
   a^2-A b\right)+\frac{1}{2} a^2 \log ^2 x
   -\frac{ab}{2}\frac{\text{L}_{A}+1+ \log x}{x}+c_{2}
   \bigg]+\mc O(\lambda^{3}).
\end{split}
\ee
where $\text{L}_A = \log(2\,A\,\lambda)+\gamma_{\rm E}$.
The higher orders  $\lambda^{n}$ are complicated but have a similar structure 

\subsubsection{Analysis of the ultraviolet region}

In the UV region, we first expand the exact equation 
\be
\label{4.23}
-\psi''(x)+(\alpha^{2}+\lambda\,V(x))\,\psi(x) = 0,
\ee
in powers of $\lambda$
\be
\label{4.24}
\psi(x) = 1+\lambda\,\psi_{1}(x)+\lambda^{2}\,\psi_{2}(x)+\lambda^{3}\,\psi_{3}(x)+\dots.
\ee
This leads to 
\be
\label{4.25}
\begin{split}
\psi_{1}''(x) &= V(x), \\
\psi_{2}''(x) &= A^{2}+V(x)\,\psi_{1}(x), \\
\psi_{3}''(x) &= A^{2}\,\psi_{1}(x)+V(x)\,\psi_{2}(x), \qquad \text{and so on.}
\end{split}
\ee
Now, we want to solve these equations with the condition $\psi_{n}'(0)=0$. The values
$\psi_{n}(0)$ are not relevant because they may be absorbed by a redefinition of $\psi(x)$ by a 
factor that depends on $\lambda$ but not on $x$. We shall choose $\psi_{n}(0)=0$. Thus, (\ref{4.25})
has to be solved with 
\be
\label{4.26}
\psi_{n}(0)=\psi_{n}'(0) = 0.
\ee
Given the solution, we want to expand it at large $x$ and compare
with (\ref{4.22}), after the educated guess, including switchback logarithmic terms,  
\be
\label{4.27}
A(\lambda) = c_{1,0}+(c_{2,1}\,\log\lambda+c_{2,0})\,\lambda+(c_{3,2}\,\log^{2}\lambda
+c_{3,1}\,\log\lambda+c_{3,0})\,\lambda^{3}+\dots.
\ee
To see how this works, let us begin with $\psi_{1}(x)$, {\em i.e.} \footnote{The double integration of (\ref{4.25})
is trivially reduced to (\ref{4.28}). Alternatively, two derivatives of (\ref{4.28}) give the first of (\ref{4.25}).}
\be
\la{4.28}
\psi_{1}(x) = \int_{0}^{x}dx'\,(x-x')\,V(x').
\ee
The large $x$ expansion is obtained as follows
\be
\la{4.29}
\begin{split}
\psi_{1}(x) &= \int_{0}^{x}dx'\,(x-x')\,\bigg[V(x')+\frac{a}{(x'+1)^{2}}\bigg]
-\int_{0}^{x}dx'\frac{a\,(x-x')}{(x'+1)^{2}} \\
&= \int_{0}^{\infty}dx'\,(x-x')\,\bigg[V(x')+\frac{a}{(x'+1)^{2}}\bigg]
-\int_{x}^{\infty}dx'\,(x-x')\,\bigg[V(x')+\frac{a}{(x'+1)^{2}}\bigg]\\
&-a\,(x-\log(x+1)).
\end{split}
\ee
Computing the first integral and expanding inside the second one, we obtain 
\be
\label{4.30}
\psi_{1}(x) = C_{1}+\frac{1}{2}\,\overline V\, x+a\,\log x-\frac{b}{2\,x}+\cdots,
\ee
with 
\be
\label{4.31}
\boxed{
C_{1} = -\int_{0}^{\infty}dx\,x\,\bigg[V(x)+\frac{a}{(x+1)^{2}}\bigg].
}
\ee
Comparing the UV expansion (\ref{4.30}) with the IR expansion (\ref{4.22}), we see that we indeed
recover the correct value of $c_{1,0}$ in agreement with the known result.
At the next order, {\em i.e.} solving for the function $\psi_{2}$, we find 
\be
\la{4.32}
\begin{split}
\psi_{2}(x) &= \frac{A^{2}x^{2}}{2}+\int_{0}^{x}dx'\,(x-x')\,V(x')\,\psi_{1}(x') \\
&= 
\frac{A^{2}x^{2}}{2}
+x\,\int_{0}^{x}dx'\,\bigg[V(x')\,\psi_{1}(x') 
+\frac{a \Vb}{2}\frac{1}{x'+1}\bigg] 
-x\,\int_{0}^{x}dx'\,\bigg[
\frac{a \Vb}{2}\frac{1}{x'+1}\bigg] \\
&
-\int_{0}^{x}dx'\,x'\,\bigg[V(x')\,\psi_{1}(x') 
+\frac{a \Vb}{2}\frac{1}{x'+1}+\frac{a C_{1}+\frac{\Vb}{2}(a+b)+a^{2}\log x'}{(x'+1)^{2}}\bigg] \\
&+\int_{0}^{x}dx'\,x'\,\bigg[
\frac{a \Vb}{2}\frac{1}{x'+1}+\frac{a C_{1}+\frac{\Vb}{2}(a+b)+a^{2}\log x'}{(x'+1)^{2}}\bigg]
\end{split}
\ee
Doing the previous split in the integrals that cannot be computed by elementary means, we obtain 
\be
\la{4.33}
\begin{split}
\psi_{2}(x) &= C_{2}+\frac{A^{2}}{2}x^{2}+x\bigg(k-\frac{a\overline V}{2}\log x+\frac{a\overline V}{2}\bigg)
+\frac{a^{2}}{2}\log^{2}x\\
&+(a^{2}+aC_{1}+\frac{b\overline V}{2})\log x
-\frac{ab(1+\log x)+b C_{1}}{2x}+\frac{b^{2}}{12x^{2}}+\cdots,
\end{split}
\ee
where \footnote{An alternative more explicit form is 
$k=\frac{\Vb}{2}\,(a-C_{1})-2\,\int_{0}^{\infty}dx\,\int_{0}^{\infty}dy\, x\, V(x)\, V(x+y)$.}
\be
\la{4.34}
\boxed{
\begin{split}
k &=  \int_{0}^{\infty} dx\, \bigg[V(x)\, \psi_{1}(x)+\frac{a\,\overline V}{2}\frac{1}{1+x}\bigg], \\
C_{2} &= -\int_{0}^{\infty}dx\,x\,\bigg[V(x)\,\psi_{1}(x) 
+\frac{a \Vb}{2}\frac{1}{x+1}+\frac{a C_{1}+\frac{\Vb}{2}(a+b)+a^{2}\log x}{(x+1)^{2}}\bigg]
-\frac{a^{2}}{6}(\pi^{2}-6)-\frac{a\Vb}{2}.
\end{split}
}
\ee
Finally, let us consider $\psi_{3}$, we need the expansion of it up to a constant because this is the last stage of our
computation. 
\be
\la{4.35}
\begin{split}
\psi_{3}'(x) &= \int_{0}^{x}dx' \bigg[A^{2}\,\psi_{1}(x')+V(x')\,\psi_{2}(x')\bigg] \\
&= A^{2} \int_{0}^{x}dx' \bigg[\psi_{1}(x')-C_{1}-\frac{\Vb}{2}x'-a \log x'+\frac{b}{2(x'+1)}\bigg]\\
&+A^{2} \int_{0}^{x}dx' \bigg[C_{1}+\frac{\Vb}{2}x'+a \log x'-\frac{b}{2(x'+1)}\bigg]\\
&+\int_{0}^{x}dx' \bigg[V(x')\,\psi_{2}(x')+\frac{a A^{2}}{2}
+\frac{-a^{2}\Vb\log x'+a^{2}\Vb+2a k+A^{2}b}{2(x'+1)}
\bigg] \\
&-\int_{0}^{x}dx' \bigg[\frac{a A^{2}}{2}
+\frac{-a^{2}\Vb\log x'+a^{2}\Vb+2a k+A^{2}b}{2(x'+1)}
\bigg].
\end{split}
\ee
Integrating,
\be
\la{4.36}
\begin{split}
\psi_{3}(x) &= C_{3}+\frac{A^{2}\Vb}{12}\,x^{3}+\frac{A^{2}}{2}(a \log x-2a+C_{1})\, x^{2}\\
&
+x\,\bigg(\frac{a^{2}\Vb}{4}\log^{2}x+(1-\tfrac{\pi^{2}}{12})\,a^{2}\Vb
-\log x(a\,(a\Vb+k)+A^{2}b)+a k+A^{2}(b+k')+k''\bigg)\\
&+\mc O(x^{0}\log^{n}x),
\end{split}
\ee
where
\be
\la{4.37}
\boxed{
\begin{split}
k' &= \int_{0}^{\infty}dx \bigg[\psi_{1}(x)-C_{1}-\frac{\Vb}{2}x-a \log x+\frac{b}{2(x+1)}\bigg],\\
k'' &= \int_{0}^{\infty}dx \bigg[V(x)\,\psi_{2}(x)+\frac{a A^{2}}{2}
+\frac{-a^{2}\Vb\log x+a^{2}\Vb+2a k+A^{2}b}{2(x+1)}
\bigg],
\end{split}
}
\ee
are new constants to be determined by quadrature.

\subsubsection{Matching UV and IR, final result}

The coefficients of the expansion (\ref{3.1}) may be found by comparing the expansions
in the IR and UV regions. After straighforward calculations, the general result reads
\be
\la{4.38}
\begin{split}
c_{1,0} &= -\tfrac{1}{2}\,\Vb, \\
c_{2,1} &= -a\,\Vb, \\
c_{2,0} &= \tfrac{1}{2}\,\Vb\,(-2\,a\,\log(-\Vb)+a\,(1-2\gamma_{\rm E})+C_{1})-k,\\
c_{3,2} &= -a^{2}\,\Vb, \\
c_{3,1} &= -\frac{b}{2}\,\Vb^{2}+a\,\Vb\,(-2\,a\,\log(-\Vb)-2\,a\,(1+\gamma_{\rm E})+C_{1})+2\,a\,k, \\
c_{3,0} &= \tfrac{1}{16}\,\Vb^{2}\,(-8\,b\,\log(-\Vb)+(5-8\gamma_{\rm E})\,b-4\,k') \\
&+\Vb\,\bigg\{\tfrac{1}{12}\bigg[(24-24\gamma_{\rm E}-12\,\gamma_{E}^{2}+\pi^{2})\,a^{2}
+6\,a\,(2\gamma_{\rm E}+1)\,C_{1}-6\,C_{1}^{2}+6\,C_{2}\bigg]\\
& -a^{2}\log^{2}(-\Vb)
+a\,(C_{1}-2(1+\gamma_{\rm E})\,a)\,\log(-\Vb)\bigg\}\\
&-2\,a\,k\,\log(-\Vb)-a\,(2\gamma_{\rm E}+1)\,k+C_{1}\,k-k''
\end{split}
\ee
It is convenient to write the asymptotic expansion (\ref{3.1}) by isolating a relevant IR scale
and making explicit the simplest coefficients
\be
\la{4.39}
\begin{split}
\sqrt{-E_{0}(\lambda)} &= -\tfrac{1}{2}\,\Vb\,\lambda\,\bigg[
1+2\,a\,\log\left(\frac{\lambda}{\lambda_\text{IR}}\right)\,\lambda \\
& + 
\bigg(2\,a^{2}\,\log^{2}\left(\frac{\lambda}{\lambda_\text{IR}}\right)+
(6\,a^{2}+b\,\Vb)
\,\log\left(\frac{\lambda}{\lambda_\text{IR}}\right)\,+\widetilde c_{3,0}\bigg)\,\lambda^{2}+\cdots\bigg].
\end{split}
\ee
with 
\be
\la{4.40}
\begin{split}
\log\lambda_\text{IR} &=  \frac{C_{1}}{2\,a}-\frac{k}{a\,\Vb}-\log(-\Vb)-\gamma_{\rm E}+\frac{1}{2},\\
\widetilde c_{3,0} &= -\frac{2\, k^2}{\Vb^2}+\frac{2\,k''-2\, a \, k}{\Vb}+\frac{-(9+\pi ^2)\, a^3
+6\, a^2\, C_{1}+3\, a\, (C_{1}^2-2\, C_{2})-6\, b\, k}{6\,a}\\
&+\frac{1}{8} \Vb\, \bigg[b \bigg(\frac{4 C_{1}}{a}-1\bigg)+4\, k'\bigg].
\end{split}
\ee
Notice that the exponentiation of the leading logs is recovered at this order. The quantities in  (\ref{4.40})
can be computed in terms of (\ref{3.12}), (\ref{4.31}), (\ref{4.34}), and (\ref{4.37}). These require quadratures
that involves the functions $\psi_{1}(x)$ and $\psi_{2}(x)$ whose expression must be computed from, see (\ref{4.28}),
\be
\la{4.41}
\psi_{1}(x) = \int_{0}^{x}dx'\,(x-x')\,V(x'), \qquad
\psi_{2}(x) = \int_{0}^{x}dx'\,(x-x')\,\bigg[A^{2}+V(x')\,\psi_{1}(x')\bigg],
\ee
and the quantity $A$ may be replaced by $-\tfrac{1}{2}\,\Vb$ at this order.

\section{Checks and applications}
\label{sec:checks}

\subsection{Check with the potential $V_{\rm I}(x) = -\frac{1}{1+x^{2}}$}

For the potential $V_{\rm I}$, we obtain the following explicit 
functions in (\ref{4.41}) ($\psi_{2}(x)$ is real for $x>0$ although this is not obvious from 
the following expression)
\be
\la{5.1}
\begin{split}
\psi_{1}(x) &=  \frac{1}{2} \log (x^2+1)-x \arctan x, \\
\psi_{2}(x) &= \frac{1}{24} (24\, i\, x\, \text{Li}_2(-e^{2 i \arctan x})+3 \log (x^2+1) (\log
   (x^2+1)+4) \\ 
   & -12\, x\, \bigg(\log
   (x^2+1)+4 \log \left(\frac{2 i}{x+i}\right)+2\bigg)
   \arctan x+\pi ^2\, x\, (3 x+2\, i)\\
   &+12\, (1+2\, i\, x) \arctan^{2} x).
\end{split}
\ee
The various constants needed in (\ref{4.40}) can be computed in closed form and read
\be
\la{5.2}
\begin{split}
\Vb &= -\pi, \ a=1,\ b=0,\ C_{1}=1,\ k=-\pi\,\log 2,\ C_{2}=3+\frac{\pi^{2}}{8}, \\
k' &= \frac{\pi}{4},\ k'' = -\frac{\pi^{3}}{8}-\pi\,\log^{2} 2-\pi\,\log 2.
\end{split}
\ee
Replacing them  in (\ref{4.38}), we match the content of Tab.~(\ref{tab1}). In the form (\ref{4.39}), it reads
\be
\la{5.3}
\begin{split}
\sqrt{-E_{0}(\lambda)} &= \frac{\pi}{2}\,\lambda\bigg[
1+2\,\log\left(\frac{\lambda}{\lambda_\text{IR}}\right)\,\lambda + 
\bigg(2\,\log^{2}\left(\frac{\lambda}{\lambda_\text{IR}}\right)+6
\,\log\left(\frac{\lambda}{\lambda_\text{IR}}\right)\,
-4 \left(2 \mathcal{L}^2+\mathcal{L}-1\right)\\
&+\frac{8
   \mathcal{L}^2+4 \mathcal{L}-7}{\pi }-\frac{\pi }{6}
   \bigg)\,\lambda^{2}+\cdots\bigg].
\end{split}
\ee
with $\log\lambda_{\rm IR} = 1-2\,\mc L$.

\subsection{Check with the potential $V_{\rm II}(x) = -\frac{1}{4\,(1+|x|)^{2}}$}

For the potential $V_{\rm II}$, we obtain 
\be
\la{5.4}
\begin{split}
\psi_{1}(x) &= \frac{1}{4} (\log (x+1)-x), \\
\psi_{2}(x) &= \frac{1}{32} ((x-6) x+\log (x+1) (2 x+\log (x+1)+6)).
\end{split}
\ee
The various constants are now
\be
\la{5.5}
\begin{split}
\Vb &= -\frac{1}{2} , \ a=\frac{1}{4},\ b=-\frac{1}{2},\ C_{1}=0,\ k=-\frac{1}{8},\ C_{2}=\frac{1}{16}, \\
k' &= \frac{1}{4},\ k'' = -\frac{13}{128}-\frac{\pi^{2}}{384}.
\end{split}
\ee
Again, replacing them  in (\ref{4.38}), we match the content of Tab. (\ref{tab1}).
In the form (\ref{4.39}), it reads
\be
\la{5.6}
\begin{split}
\sqrt{-E_{0}(\lambda)} &= \frac{1}{4}\,\lambda\bigg[
1+\frac{1}{2}\,\log\left(\frac{\lambda}{\lambda_\text{IR}}\right)\,\lambda + 
\bigg(\frac{1}{8}\,\log^{2}\left(\frac{\lambda}{\lambda_\text{IR}}\right)+\frac{5}{8}
\,\log\left(\frac{\lambda}{\lambda_\text{IR}}\right)\,-\frac{11}{32}
   \bigg)\,\lambda^{2}+\cdots\bigg].
\end{split}
\ee
with $\log\lambda_{\rm IR} = \log 2-\gamma_{\rm E}-\tfrac{1}{2}$.

\subsection{Expansion for new non-soluble potentials, examples}

As an application of our method to a new potentials that are not soluble, nor already treated by other methods, 
we consider the following two cases
\be
\la{5.7}
V_{\rm III}(x) = -\frac{|x|}{4\,(1+|x|)^{3}}, \qquad V_{\rm IV}(x) = -\frac{x^{2}}{(1+x^{2})^{2}}.
\ee
For the potential $V_{\rm III}$, we obtain 
\be
\la{5.8}
\begin{split}
\psi_{1}(x) &= \frac{1}{4} \log (x+1)-\frac{x (x+2)}{8 (x+1)},\\
\psi_{2}(x) &= \frac{1}{384 (x+1)^2}\bigg(3 x^4-38 x^3+12 x^3 \log (x+1)-99 x^2+12 x^2 \log ^2(x+1)\\
&+72
   x^2 \log (x+1)-60 x+24 x \log ^2(x+1)+12 \log ^2(x+1)\\
   &+120 x
   \log (x+1)+60 \log (x+1)\bigg).
\end{split}
\ee
The various constants are
\be
\la{5.9}
\begin{split}
\Vb &= -\frac{1}{4}, \ a=\frac{1}{4},\ b=-\frac{3}{4},\ C_{1}=-\frac{1}{8},\ k=-\frac{1}{12},
\ C_{2}=-\frac{1}{192}, \\
k' &= \frac{1}{4},\ k'' = -\frac{319}{4608}-\frac{\pi^{2}}{768}.
\end{split}
\ee
Replacing them  in (\ref{4.38}), we obtain the new asymptotic expansion
\be
\la{5.10}
\begin{split}
\sqrt{-E_{0}(\lambda)} &= \frac{1}{8}\,\lambda\bigg[
1+\frac{1}{2}\,\log\left(\frac{\lambda}{\lambda_\text{IR}}\right)\,\lambda + 
\bigg(\frac{1}{8}\,\log^{2}\left(\frac{\lambda}{\lambda_\text{IR}}\right)+\frac{9}{16}
\,\log\left(\frac{\lambda}{\lambda_\text{IR}}\right)\,-\frac{43}{144}
   \bigg)\,\lambda^{2}+\cdots\bigg].
\end{split}
\ee
with $\log\lambda_{\rm IR} = 2\log 2-\gamma_{\rm E}-\tfrac{13}{12}$. 
A similar calculation may be performed for the potential $V_{\rm IV}$. We omit the details and just write down the 
final expansion
\be
\la{5.11}
\begin{split}
\sqrt{-E_{0}(\lambda)} &= \frac{\pi}{4}\,\lambda\bigg[
1+2\,\log\left(\frac{\lambda}{\lambda_\text{IR}}\right)\,\lambda + 
\bigg(2\,\log^{2}\left(\frac{\lambda}{\lambda_\text{IR}}\right)+6
\,\log\left(\frac{\lambda}{\lambda_\text{IR}}\right)\\
&
-\frac{17\,\pi^{2}}{48}-\frac{121}{32}-2\,\log^{2}2-\frac{7}{2}\log 2
   \bigg)\,\lambda^{2}+\cdots\bigg].
\end{split}
\ee
with $\log\lambda_{\rm IR} =-\gamma_{\rm E}+\frac{3}{8}-\log\pi$. 

\subsection{Simple  extensions}

We remark that our expressions are sufficiently simple to deal with potentials that depend on 
parameters. One neat example is the potential 
\be
\la{5.12}
V(x) = -\frac{\beta-1+|x|}{4\,(1+|x|)^{3}}, \qquad \beta\ge 0.
\ee
For this potential we obtain the following generalization of the $\beta=1$ case in (\ref{5.10}), 
\be
\la{5.13}
\begin{split}
\sqrt{-E_{0}(\lambda)} &= \frac{\beta}{8}\,\lambda\bigg[
1+\frac{1}{2}\,\log\left(\frac{\lambda}{\lambda_\text{IR}}\right)\,\lambda + 
\bigg(\frac{1}{8}\,\log^{2}\left(\frac{\lambda}{\lambda_\text{IR}}\right)+
\frac{6+4\,\beta-\beta^{2}}{16}
\,\log\left(\frac{\lambda}{\lambda_\text{IR}}\right)\\
&
-\frac{1}{72 \beta ^2}+\frac{5}{144 \beta
   }-\frac{13}{144}-\frac{217 \beta }{576}+\frac{25 \beta
   ^2}{144}-\frac{5 \beta ^3}{192}
   \bigg)\,\lambda^{2}+\cdots\bigg].
\end{split}
\ee
with $\log\lambda_{\rm IR} =2\log 2-\gamma_{\rm E}-\log\beta-\frac{1}{3\,\beta}-\frac{7}{6}+\frac{5}{12}\,\beta$. 
In general, it is clear that if the expansion may be worked out in details for $V_{1}$ and $V_{2}$, then it may be 
computed for any linear combination of them. Also, the class of potentials may be enlarged. One interesting example is
that of a potential with asymptotic form at infinity including a term $\sim \log|x|/x^{4}$. Just to give a simple 
example, we studied the case
\be
\la{5.14}
V(x) = -\frac{1}{(1+|x|)^{2}}+\frac{9}{2}\,\frac{\log(1+|x|)}{(1+|x|)^{4}},
\ee
where the relative weight of the first and second terms has been chosen to have $\Vb=-1$. We find in this case,
\be
\la{5.15}
\begin{split}
\sqrt{-E_{0}(\lambda)} &= \frac{1}{2}\,\lambda\bigg[
1+2\,\log\left(\frac{\lambda}{\lambda_\text{IR}}\right)\,\lambda + 
\bigg(2\,\log^{2}\left(\frac{\lambda}{\lambda_\text{IR}}\right)+
8
\,\log\left(\frac{\lambda}{\lambda_\text{IR}}\right)
-\frac{141990237413}{2^{6}\,3^{2}\,5^{6}\,7^{4}}
   \bigg)\,\lambda^{2}+\cdots\bigg],
\end{split}
\ee
with $\log\lambda_{\rm IR} = -\gamma_{\rm E}-\frac{3313}{3000}$.

\subsection{Numerical tests}
\label{subsec:numerical}

We present a numerical test of the accuracy of the asymptotic expansions (\ref{5.3}), (\ref{5.6}), 
and the new (\ref{5.10},\ref{5.11}). The convergence radius of these expansions is expected to be zero, but looking at sufficiently
small $\lambda$, it is possible to appreciate the improvement associated with the higher order terms. The
following tables will present the first, second, and third order expansion of $\sqrt{-E_{0}(\lambda)}$ for the
four considered potentials. The last column, labeled {\em exact} is obtained by finding the ground state 
in the interval $x\in[0,L]$.~\footnote{Of course, for $V_{\rm II}$ we solve (\ref{2.2}).}
For sufficiently large $L$, the number of stable digits in the result doubles as $L$ is doubled
showing exponential convergence at large $L$. Our results at $\lambda=10^{-1}, 10^{-2}, 10^{-3}$ (
and $5\cdot 10^{-4}$ for $V_{\rm IV}$) are
\be
\label{5.16}
\begin{split}
& \hskip 3.5cm V_{\rm I}(x) = -\frac{1}{1+x^{2}}, \\
&
\begin{array}{|c|c|c|c|c|}
\hline
\lambda &  1^{\text{st}} & 2^{\text{nd}} & 3^{\text{rd}} & \text{exact}  \\
\hline
10^{-1} &  0.1570796327 & 0.1291982378 & 0.1076056359 & 0.1169558258 \\
10^{-2} &  0.0157079633 & 0.0147057709 & 0.0146919732 & 0.0147016491 \\
10^{-3} &  0.0015707963 & 0.0015535406 & 0.0015535679 & 0.0015535766 \\
\hline
\end{array}
\end{split}
\ee
\be
\label{5.17}
\begin{split}
& \hskip 3cm V_{\rm II}(x) = -\frac{1}{4\,(1+|x|)^{2}}, \\
& \begin{array}{|c|c|c|c|c|}
\hline
\lambda &  1^{\text{st}} & 2^{\text{nd}} & 3^{\text{rd}} & \text{exact}  \\
\hline
10^{-1} &   0.025  &  0.0226018542 & 0.0223311706 & 0.0223479210 \\
10^{-2} &  0.0025  & 0.0024472362 & 0.0024470475 & 0.0024470589 \\
10^{-3} & 0.00025 & 0.0002491845 & 0.0002491848 & 0.0002491848 \\
\hline
\end{array}
\end{split}
\ee
\be
\label{5.18}
\begin{split}
& \hskip 3cm V_{\rm III}(x) = -\frac{|x|}{4\,(1+|x|)^{3}}, \\
& \begin{array}{|c|c|c|c|c|}
\hline
\lambda &  1^{\text{st}} & 2^{\text{nd}} & 3^{\text{rd}} & \text{exact}  \\
\hline
10^{-1} & 0.0125     & 0.0112322935 &  0.0111166333 & 0.0111254679 \\
10^{-2} & 0.00125   & 0.0012229318 &  0.0012228830 & 0.0012228880   \\
10^{-3} & 0.000125 & 0.0001245854 &  0.0001245856 & 0.0001245856\\
\hline
\end{array}
\end{split}
\ee
\be
\label{5.19}
\begin{split}
& \hskip 3cm V_{\rm IV}(x) = -\frac{x^{2}}{(1+x^{2})^{2}}, \\
& \begin{array}{|c|c|c|c|c|}
\hline
\lambda &  1^{\text{st}} & 2^{\text{nd}} & 3^{\text{rd}} & \text{exact}  \\
\hline
10^{-1} & 0.0785398163 & 0.0635286655 & 0.0520846293 & 0.0578934148 \\
10^{-2} & 0.0078539816 & 0.0073421809 & 0.0073351273 & 0.0073412803 \\
10^{-3} & 0.0007853982 & 0.0007766633 & 0.0007766773 & 0.0007766830 \\
5\cdot 10^{-4} 
              & 0.0003926991 & 0.0003902432 & 0.0003902461 & 0.0003902468 \\
\hline
\end{array}
\end{split}
\ee
From the above tables, it is possible to check that the third order gives a sensible improvement at these small 
values of $\lambda$. The pattern is the same for all the considered potentials and, in particular, confirms
the new expansion (\ref{5.10}). Just as a final example, we also report the numerical results associated
with the potential (\ref{5.14}) 
\be
\label{5.20}
\begin{split}
& \hskip 2cm V(x) = -\frac{1}{(1+|x|)^{2}}+\frac{9}{2}\,\frac{\log(1+|x|)}{(1+|x|)^{4}}, \\
& \begin{array}{|c|c|c|c|c|}
\hline
\lambda &  1^{\text{st}} & 2^{\text{nd}} & 3^{\text{rd}} & \text{exact}  \\
\hline
10^{-1} & 0.05       & 0.0437896391 &  0.0384057384 & 0.0384464983 \\
10^{-2} & 0.005     & 0.0047076379 &  0.0047012055 & 0.0047018683 \\
10^{-3} & 0.0005   & 0.0004947738 &  0.0004947769 & 0.0004947770\\
\hline
\end{array}
\end{split}
\ee

\section{A further consistency check  from $\mc N=6$ ABJ(M) theory}
\label{sec:susy}

Apart from general questions, our initial motivation came from the Quantum Spectral Curve results on the static 
quark-antiquark potential in $\mc N=4$ super Yang-Mills gauge theory.
Quite recently, new interesting results appeared in the study of the cusp anomalous dimension in $\mc N=6$ 
ABJ(M) theory. A scaling limit has been identified where ladder diagrams dominate and the 
Bethe-Salpeter equation leads again to a Schr\"odinger problem 
\cite{Bonini:2016fnc}. This problem is studied
for generic Wilson loop Euclidean cusp angle $\varphi$. The anti-parallel limit $\varphi\to \pi$ is 
the one relevant for the calculation of the static potential, but requires some care as it is singular as in $\mc N=4$ SYM. 
On general grounds, it is expected that the larger amount of supersymmetry should be 
responsible for cancellation of the non-analytic logarithms \cite{Griguolo:2012iq}.
Technically, in the ladder approximation, this feature is due to the fact that the 
Schr\"odinger potential in the ladder approximation is supersymmetric, {\em i.e.} it has the 
the structure of 
Witten's supersymmetric quantum mechanics \cite{Witten:1981nf}. In particular, the $\mc N=6$ ABJ(M)
version of the potential (\ref{1.6}) -- relevant for $\mc N=4$ SYM -- reads \footnote{
We thanks the authors of \cite{Griguolo:2012iq,Bonini:2016fnc}, and in particular D. Seminara, for 
providing the details of the $\varphi\to \pi$ limit in $\mc N=6$ 
ABJ(M) theory.
}
\be
\la{6.1}
H = -\frac{d^{2}}{dx^{2}}-\frac{\lambda^{2}}{x^{2}+1}-\frac{\lambda}{(x^{2}+1)^{3/2}}, \qquad \lambda>0.
\ee
The exact bound state wave function is $\psi(x) = \mathcal N\, \exp(-\lambda\,\sqrt{x^{2}+1})$, with binding 
energy $E_{0} = -\lambda^{2}$, without (logarithmic) corrections. We can put this problem in the form (\ref{1.1}), by considering the potential
\be
\label{6.2}
V(x) = -\frac{\mu}{x^{2}+1}-\frac{1}{(x^{2}+1)^{3/2}},
\ee
where $\mu$ has to be replaced by $\lambda$ at the end of the calculation. This is a well-defined procedure 
and the replacement 
$\mu\to \lambda$ simply induces a rearrangement of the asymptotic expansion. Applying (\ref{4.39}) to (\ref{6.2}),
we obtain the non-trivial expansion at generic fixed $\mu$
\be
\label{6.3}
\begin{split}
\sqrt{-E_{0}(\lambda; \mu)} &= \frac{2+\pi\,\mu}{2}\,\lambda\,\bigg[
1+2\,\mu\,\log\left(\frac{\lambda}{\lambda_\text{IR}(\mu)}\right)\,\lambda
+\bigg(
2\,\mu^{2}\,\log^{2}\left(\frac{\lambda}{\lambda_\text{IR}(\mu)}\right)\\
&
-(2+\pi\,\mu-6\,\mu^{2})\,\log\left(\frac{\lambda}{\lambda_\text{IR}(\mu)}\right)
-\frac{\pi}{2\,\mu}-\frac{\pi^{2}}{8}+\mc O(\mu)
\bigg)\,\lambda^{2}
\bigg]+\mc O(\lambda^{4}),
\end{split}
\ee
where
\be
\label{6.4}
\log\lambda_{\rm IR}(\mu) = \frac{\pi+4\,\mu+2\,\pi\,\mu^{2}}{4\,\mu+2\,\pi\,\mu^{2}}
+\frac{2-\pi\,\mu}{2+\pi\,\mu}\,\log 2-\log(2+\pi\,\mu)-\gamma_{\rm E}.
\ee
Replacing $\mu = \lambda$, the terms in (\ref{6.3}) and  (\ref{6.4}) rearrange nicely to give
\be
\label{6.5}
\sqrt{-E_{0}(\lambda; \lambda)} = \lambda + \mc O(\lambda^{4}),
\ee
in agreement with the exact binding energy $E_{0} = -\lambda^{2}$. This is a further non-trivial 
check of our  expansion (\ref{4.39}).

\section{Universal resummation of leading infrared logarithms}
\label{sec:resummation}

The resummation results in (\ref{1.8}) and (\ref{2.4}) are consistent with (\ref{4.39}), (\ref{4.40}).
It is tempting to conjecture that leading logarithms indeed exponentiate at all orders, {\em i.e.} that 
the expression
\be
\la{7.1}
\left. \sqrt{-E_{0}(\lambda)} \right|_{\rm LO} = -\frac{\Vb}{2}\,\lambda\,e^{2\,a\,\lambda\log\lambda},
\ee
captures all terms of the form $\lambda^{n+1}\,\log^{n}\lambda$. Exponentiation results like (\ref{7.1})
appears in applications of matched asymptotic expansions often together with Renormalization Group 
techniques \cite{ward1993summing,titcombe1999summing,deville2008analysis,lagree2010asymptotic}.
However, a general treatment is missing and we devote this section to the analysis of (\ref{7.1}) in our 
specific setup. A physical argument for (\ref{7.1}) is that the non-analytic logarithms of the coupling 
are an infrared effect related to the long-distance behaviour of the potential. This somewhat explains
why resummation involves essentially the coefficient $a$ in (\ref{1.4}). Nevertheless, it is amusing that some
global information about the potential is needed in the form of the prefactor in (\ref{7.1}) that depends on $\Vb$.
An interesting further check of (\ref{7.1}) may be presented by considering a soluble
cut-off version of the potential
$V_{\rm II}(x)$, as discussed in App.~(\ref{app:cutoff}).

\medskip
A general proof of (\ref{7.1}) may be given exploiting the clever splitting of the Birman-Schwinger operator (\ref{3.2})
introduced in \cite{avron1981}. To this aim, we define the operators $P_{\alpha}$, $Q_{\alpha}$ with integral
kernels
\be
\la{7.2}
\begin{split}
P_{\alpha}(x,y) &= -\frac{1}{2}\, |V(x)|^{1/2}\,e^{-\alpha\,|x|}\,e^{-\alpha\,|y|}\,|V(y)|^{1/2}\, \text{sign}(V(x)), \\
Q_{\alpha}(x,y) &= -\frac{1}{2}\, |V(x)|^{1/2}\,\bigg[
e^{-\alpha\,|x-y|}-e^{-\alpha\,|x|}\,e^{-\alpha\,|y|}\bigg]\,|V(y)|^{1/2}\, \text{sign}(V(x)).
\end{split}
\ee
According to the Birman-Schwinger principle \cite{reed1978analysis}, the Schr\"odinger operator in (\ref{1.1})
has the eigenvalue $-\alpha^{2}$ if and only if $P_{\alpha}+Q_{\alpha}$ has eigenvalue $\alpha/\lambda$ or, 
equivalently, if $1-\lambda\,\alpha^{-1}\,(P_{\alpha}+Q_{\alpha})$ is not invertible -- compare with (\ref{3.1}).
Since $P_{\alpha}$ is rank 1, this translates into the condition 
\be
\la{7.3}
\begin{split}
\alpha &= \lambda\,\text{Tr}\bigg(P_{\alpha}\, (1-\lambda\,\alpha^{-1}\,Q_{\alpha})^{-1}\bigg) \\
&= \lambda\,\text{Tr}(P_{\alpha})+\frac{\lambda^{2}}{\alpha}\,\text{Tr}(P_{\alpha}\, Q_{\alpha})
+\frac{\lambda^{3}}{\alpha^{2}}\,\text{Tr}(P_{\alpha}\, Q_{\alpha}^{2})+\cdots.
\end{split}
\ee
The first term is 
\be
\la{7.4}
\text{Tr}(P_{\alpha}) = -\frac{1}{2}\,\int_{\mathbb R} dx\, V(x)\, e^{-2\alpha\, |x|} = 
-\frac{1}{2}\,\Vb+\int_{0}^{\infty}dx\, V(x)\, (1-e^{-2\,\alpha\, x}).
\ee
The second term can be computed by adding and subtracting the leading large distance term 
$-a/(1+x)^{2}$ to $V(x)$ and we obtain 
\be
\la{7.5}
\text{Tr}(P_{\alpha}) =  
-\frac{1}{2}\,\Vb+2\,a\,\alpha\,\log\alpha+\text{subleading terms}.
\ee
Let us analyze the next terms
\be
\la{7.6}
\begin{split}
\text{Tr}& (P_{\alpha}\,Q_{\alpha}^{n-1}) = \frac{(-1)^{n}}{2^{n}}
\int_{\mathbb R} dx_{1}\cdots dx_{n}\, V(x_{1})\cdots V(x_{n}) \,
e^{-\alpha\,|x_{1}|-\alpha\,|x_{2}|}\\
&\bigg[
e^{-\alpha\,|x_{2}-x_{3}|}-e^{-\alpha\,|x_{2}|}\,e^{-\alpha\,|x_{3}|}\bigg]\cdots
\bigg[
e^{-\alpha\,|x_{n}-x_{1}|}-e^{-\alpha\,|x_{n}|}\,e^{-\alpha\,|x_{1}|}\bigg] \\
&= \frac{(-1)^{n}}{2^{n-1}}
\int_{x_{i}\ge 0} dx_{1}\cdots dx_{n}\, V(x_{1})\cdots V(x_{n}) \,
e^{-\alpha\,x_{1}-\alpha\,x_{2}}\\
& \bigg[
e^{-\alpha\,|x_{2}-x_{3}|}+e^{-\alpha\,(x_{2}+x_{3})}-2\,e^{-\alpha\,x_{2}}\,e^{-\alpha\,x_{3}}\bigg]
 \cdots\bigg[
e^{-\alpha\,|x_{n}-x_{1}|}+e^{-\alpha\,(x_{n}+x_{1})}-2\,e^{-\alpha\,x_{n}}\,e^{-\alpha\,x_{1}}\bigg] \\
&= \frac{(-1)^{n}}{2^{n-1}}
\int_{x_{i}\ge 0} dx_{1}\cdots dx_{n}\, V(x_{1})\cdots V(x_{n}) \,
e^{-\alpha\,x_{1}-\alpha\,x_{2}}\\
& \bigg[
e^{-\alpha\,|x_{2}-x_{3}|}-e^{-\alpha\,x_{2}}\,e^{-\alpha\,x_{3}}\bigg]
 \cdots\bigg[
e^{-\alpha\,|x_{n}-x_{1}|}-e^{-\alpha\,x_{n}}\,e^{-\alpha\,x_{1}}\bigg] \\
\end{split}
\ee
The most singular contribution comes from those permutations of $\{x_{k}\}$ with 
a specific maximal ordering with all large terms $|x_{i}-x_{j}|$. \footnote{
Alternatively, one may replace at this point $V$ by its leading asymptotic and reduce to the problem in Sec.~(\ref{sec:solvable}) 
where computations may be done more explicitly.}
\be
\la{7.7}
\begin{split}
\text{Tr}& (P_{\alpha}\,Q_{\alpha}^{n-1})_{\rm LO} =  \frac{(-1)^{n}}{2^{n-1}}\times\frac{2^{n-1}}{n!}
\int_{0\le x_{1}\le \cdots \le x_{n}}\, V(x_{1})\cdots V(x_{n}) e^{-2\alpha x_{n}}\prod_{k=1}^{n-1}
(1-e^{-2\alpha x_{k}})
\end{split}
\ee
Evaluating the leading logarithmic term integrating first in $x_{n}\in [x_{n-1},\infty]$ and so on, one obtains
\be
\la{7.8}
\text{Tr} (P_{\alpha}\,Q_{\alpha}^{n-1}) = \frac{(-1)^{n}}{n!}\,[-(2\,a\,\alpha\,\log\alpha)^{n}],
\ee
and therefore (\ref{7.3}) reads at leading logarithmic order 
\be
\la{7.9}
\alpha = -\frac{\Vb}{2}\,\lambda+\sum_{k=1}^{\infty} \frac{(-1)^{n+1}\,2^{n}}{n!}\,\frac{\lambda^{n}}{\alpha^{n-1}}
(a\,\alpha\,\log\alpha)^{n}.
\ee
The series sums to $\alpha-\alpha^{1-2\,a\,\lambda}$. Solving for $\alpha$ and manipulating up to subleading logarithmic terms $\lambda^{n}\log^{m}\lambda$ with 
$n>m+1$, we get
\be
\la{7.10}
\alpha = \left(-\frac{\Vb}{2}\,\lambda\right)^{\frac{1}{1-2a\lambda}} \simeq
-\frac{\Vb}{2}\,e^{\frac{\log\lambda}{1-2\,a\,\lambda}} \simeq
-\frac{\Vb}{2}\,e^{(1+2\,a\,\lambda)\,\log\lambda} =  -\frac{\Vb}{2}\,\lambda\,e^{2\,a\,\lambda\log\lambda},
\ee
that is (\ref{7.1}).

\section{Conclusions}
\label{sec:conclusions}

In summary, we have reconsidered the one-dimensional Schr\"odinger problem (\ref{1.1}) for long-range 
potentials in the class (\ref{1.4}) with the aim
of deriving an improved asymptotic expansion of the binding energy of the (unique) bound state 
that is present at weak coupling. Non-analytic terms are present at all orders in the coupling and are of course
of infrared origin. Our setup is the simplest possible where such infrared problems occur and it is not possible to 
apply the powerful machinery of relativistic quantum field theory. In particular, there are no simple cut-off 
procedures that capture them accurately fixing the correct long-distance scales. We computed the general 
third order asymptotic expansion by methods coming from boundary layer theory. In principle, similar
techniques may be applied to more difficult problems involving different behaviours at infinity, like 
for instance $V(x)\sim \log(x)/x^{2}$ or even $\log(x)/x^{3}$, where the extension is non trivial.
It would be quite interesting to explore similar problems in more than one dimension.

\section*{Acknowledgments}
We  thank  D. Seminara,  G. Macorini and L. Girlanda for discussions and comments.

\appendix

\section{All order resummation of LO and NLO  infrared logarithms for the potential $V_{\rm II}$}
\label{App:resum}

The quantization condition (\ref{2.2}) may be written 
\be
\la{A.1}
(1-2\,\nu)\,\text{K}_{\nu}(\alpha)-2\,\alpha\,\text{K}_{\nu-1}(\alpha)=0,\qquad \nu = \frac{1}{2}\sqrt{1-\lambda}.
\ee
We multiply it  by $\alpha^{\nu}$ and expand at $\alpha\to 0$. After some simple manipulation we obtain
\be
\la{A.2}
\alpha = \left(\frac{4^{\nu}\,(2\,\nu-1)\,\Gamma(\nu)}{(2\,\nu+1)\,\Gamma(-\nu)}\right)^{\frac{1}{2\nu}}
\bigg[1+\frac{13-4\,\nu^{2}}{4\,(4\,\nu^{4}-5\nu^{2}+1)}\,\alpha^{2}+\mc O(\alpha^{4})\bigg].
\ee
For $\lambda\to 0$, we have $\nu\to 1/2$ and $\alpha = \mc O(\lambda)$. Thus, (\ref{A.2}) may be consistently 
solved iteratively $\alpha=\alpha^{(0)}+\alpha^{(1)}+\cdots$, {\em i.e.}
\be
\la{A.3}
\begin{split}
\alpha^{(0)} = \left(\frac{4^{\nu}\,(2\,\nu-1)\,\Gamma(\nu)}{(2\,\nu+1)\,\Gamma(-\nu)}\right)^{\frac{1}{2\nu}}, 
\alpha^{(1)} = \frac{13-4\,\nu^{2}}{4\,(4\,\nu^{4}-5\nu^{2}+1)}
\,\left(\frac{4^{\nu}\,(2\,\nu-1)\,\Gamma(\nu)}{(2\,\nu+1)\,\Gamma(-\nu)}\right)^{\frac{1}{\nu}}.
\end{split}
\ee
We can expand these expressions for $\lambda\to 0$ and introduce the 
quantity $\text{L} = \log\lambda+\gamma_{\rm E}-\log 2
+\tfrac{1}{2}$.
The terms in (\ref{A.3}) of the form $\lambda^{n+1}\,\text{L}^{n}$ and $\lambda^{n+2}\,\text{L}^{n}$ are
\be
\la{A.4}
\begin{split}
\alpha^{(0)} &=  \tfrac{1}{4}\,\lambda\,
e^{\frac{1}{2}\,\lambda\,\text{L}}\,\bigg(
1-\tfrac{1}{4}\,\lambda+\tfrac{3}{8}\,\lambda^{2}\,\text{L}
\bigg)
+\text{N${}^{2}$LO}, \\
\alpha^{(1)} &=  \tfrac{1}{16}\,\lambda^{2}\,e^{\frac{3}{2}\,\lambda\,\text{L}}
+\text{N${}^{2}$LO}.
\end{split}
\ee
The third and higher corrections do not contribute. Summing the terms in (\ref{A.4}) we
 prove the quoted result (\ref{2.4}).

\section{Euristic derivation of the leading infrared logarithm at third order}
\label{App:3naive}

The following combination in  (\ref{3.3})
\be
\label{B.1}
(\text{tr}K_{\alpha})^{2}-\text{tr}K_{\alpha}^{2},
\ee
is essentially $F(\alpha)$ in (\ref{3.5}). For this quantity we know that its small $\alpha\to 0^{+}$ expansion 
has leading term $\alpha\,(\#\log\alpha+\#)$. \footnote{Here and in the following we shall denote by $\#$
numerical coefficients.} There are further correction $\mc O(\alpha^{2})$, but an analysis of (\ref{3.5})
shows that they do not contain $\log^{2}\alpha$ enhancement factors. This means that the $\lambda^{3}\,\log^{2}\lambda$  term comes entirely from the very last term of (\ref{3.3}), {\em i.e.}
\be
\label{B.2}
\frac{\lambda^{3}}{3}\,\text{tr}\,K_{\alpha}^{3}.
\ee
The general expression of the trace is 
\be
\label{B.3}
\begin{split}
\text{tr}\,K_{\alpha}^{3} &= \frac{1}{8\,\alpha^{3}}\int_{\mathbb R} dx\,dy\,dz V(x)\,V(y)\,V(z)\,
e^{-\alpha\,(|x-y|+|x-z|+|y-z|)} \\
&= \frac{1}{8}\,\bigg(\frac{2}{\pi}\bigg)^{3/2}\,\int_{\mathbb R}\,d\omega_{1}\,d\omega_{2}\,d\omega_{3}\,\frac{\widetilde V(\omega_{1}-\omega_{2})\,\widetilde V(\omega_{1}-\omega_{3})\,\widetilde V(\omega_{2}-\omega_{3})}{(\alpha^{2}+\omega_{1}^{2})(\alpha^{2}+\omega_{2}^{2})(\alpha^{2}+\omega_{3}^{2})}.
\end{split}
\ee
The function $\widetilde V(\omega)$ is in general a function of $|\omega|$. It is thus convenient to exploit
the symmetry of the integrand of (\ref{B.3}) with respect to permutations and write 
\be
\label{B.4}
\text{tr}\,K_{\alpha}^{3} = \frac{3}{4}\,\,\bigg(\frac{2}{\pi}\bigg)^{3/2}\,\int_{\omega_{1}<\omega_{2}<\omega_{3}}\,d\omega_{1}\,d\omega_{2}\,d\omega_{3}\,\frac{\widetilde V(\omega_{1}-\omega_{2})\,\widetilde V(\omega_{1}-\omega_{3})\,\widetilde V(\omega_{2}-\omega_{3})}{(\alpha^{2}+\omega_{1}^{2})(\alpha^{2}+\omega_{2}^{2})(\alpha^{2}+\omega_{3}^{2})}.
\ee
Finally, for $\omega>0$, we write $\widetilde V(\omega) = \widetilde V(0) + \omega\,\widetilde V'(0^{+})+\cdots$
and keep the term proportional to $\widetilde V(0)\,(\widetilde V'(0^{+}))^{2}$. We heuristically 
claim that this is the desired 
contribution containing the $\log^{2}\alpha$ term. Using again $\widetilde V(0) = \frac{1}{\sqrt{2\pi}}\overline V$
as well as (\ref{3.16}), we reduce the calculation to 
\be
\label{B.5}
\left. \text{tr}\,K_{\alpha}^{3}\right|_{*} = \frac{3\,a^{2}\,\overline V}{4\,\pi}\,\int_{\omega_{1}<\omega_{2}<\omega_{3}}\,d\omega_{1}\,d\omega_{2}\,d\omega_{3}\,\frac{
\omega _1^2+\omega _1 \omega _2-\omega _2^2-3 \omega _1 \omega _3+\omega _2
   \omega _3+\omega _3^2
}{(\alpha^{2}+\omega_{1}^{2})(\alpha^{2}+\omega_{2}^{2})(\alpha^{2}+\omega_{3}^{2})}.
\ee
where we denoted by $\left. (\cdots)\right|_{*}$ the fact that we are picking up a particular piece of $\text{tr}K^{3}_{\alpha}$. We now split
\be
\label{B.6}
\left. \text{tr}\,K_{\alpha}^{3}\right|_{*} = \frac{3\,a^{2}\,\overline V}{4\,\pi}\,\sum_{n=1}^{6} I_{n},
\ee
where the 6 integrals $I_{n}$ are associated to the 6 terms in the numerator of the r.h.s. of (\ref{B.5}). All of them 
may be computed exactly after introducing a constraint $|\omega_{i}|<1$ that 
gives convergence at large $\omega$ without changing the singular parts we are looking for. 
With some effort the partial results are :
\be
\label{B.7}
\begin{split}
I_{1} &= \frac{\pi ^2}{2 \alpha ^2}+\frac{\pi  \log \alpha -\frac{\pi ^3}{6}-2 \pi +\pi  \log
   2}{\alpha }+\pi ^2+\mc O(\alpha), \\
I_{2} &=\frac{12 \pi  \log ^2\alpha +24 \pi  \log 2 \log \alpha +\pi ^3+12 \pi  \log
   ^2 2}{12 \alpha }-4+\mc{O}(\alpha), \\
I_{3} &= \frac{2 \pi  \log \alpha +\frac{\pi ^3}{6}+2 \pi  \log 2}{\alpha }+\left(-4-\pi
   ^2\right)+\mc O(\alpha), \\
I_{4} &= \frac{12 \pi  \log ^2\alpha +24 \pi  \log 2 \log \alpha +\pi ^3+12 \pi  \log
   ^2 2}{4 \alpha }-12+\mc O(\alpha),\\
I_{5} &= -2+\mc O(\alpha), \\
I_{6} &= \frac{\pi ^2}{2 \alpha ^2}+\frac{\pi  \log \alpha -\frac{\pi ^3}{6}-2 \pi +\pi  \log
   2}{\alpha }+\pi ^2+\mc O(\alpha).
\end{split}
\ee
Summing up,
\be
\label{B.8}
\begin{split}
\sum_{n=1}^{6}\,I_{n} &= \frac{\pi ^2}{\alpha ^2}
+\frac{\pi  \left(24 \log ^2\alpha +24 (1+2 \log 2) \log
   (\alpha )+\pi ^2+24 \left(-1+\log ^2 2+\log  2\right)\right)}{6 \alpha }\\
   &+\left(\pi
   ^2-22\right)+\mc O(\alpha).
   \end{split}
\ee
The term $\sim\alpha^{-2}$ is going to cancel against other terms, see (\ref{3.3}), while
the relevant piece in  (\ref{B.2}) turns out to be 
\be
\frac{\lambda^{3}}{3}\,\text{tr} K_{\alpha}^{3} = a^{2}\,\overline V\,\frac{\lambda^{3}}{\alpha}\,
\log^{2}\alpha+\cdots,
\ee
where dots stand for terms we are not interested in. We conclude that the leading logarithmic part of 
$\alpha(\lambda)$ is obtained by inverting the expansion, (see (\ref{3.3}), (\ref{3.11}), etc.)
\be
\label{B.10}
0=1+\frac{\lambda}{2\alpha}\,\overline V + \lambda^{2}\,a\,\overline V\,\log\alpha
+\lambda^{3}\,a^{2}\,\overline V\,\log^{2}\alpha + \cdots,
\ee
that gives 
\be
\label{B.11}
\alpha = -\frac{\lambda}{2}\,\overline V\,\bigg(
1+2\,a\,\lambda\,\log\lambda+2\,a^{2}\,\lambda\,\log^{2}\lambda+\cdots
\bigg),
\ee
or, going back to (\ref{3.1}),
\be
\label{B.12}
{
c_{3,2} = -a^{2}\,\overline V,
}
\ee
in agreement with table Tab.~(\ref{tab1}). 

\section{Resummation of leading logarithms in a cut-off version of $V_{\rm II}(x)$}
\label{app:cutoff}

To support the universal exponentiation of the leading logarithms
as well as to 
show how it cannot be trivially extended to next-to-leading logarithms, we consider the 
illustrative problem associated with the following potential
\be
\la{C.1}
V(x) = 
\begin{cases}
-\frac{1}{4\,(1+x_{0})^{2}} &\text{if  $|x|\le x_{0}, \qquad x_{0}>0$,}\\
-\frac{1}{4\,(1+|x|)^{2}} &\text{if $|x|>x_{0}$.}
\end{cases}
\ee
This is a simple  modification of $V_{\rm II}(x)$ that replaces it by the constant $V_{\rm II}(x_{0})$ for 
$|x|\le x_{0}$. Thus, the asymptotic behaviour at infinity is unchanged. The study of this example is useful to 
see how the global shape of $V$ affects the asymptotic expansion of its ground state energy. The exact ground state
wave function is (with irrelevant arbitrary normalization)
\be
\la{C.2}
\psi_{0}(x) = 
\begin{cases}
\cos\left(x\,\frac{\sqrt{\lambda-4\,\alpha^{2}\,(x_{0}+1)^{2}}}{2\,(x_{0}+1)}\right) &\text{if  $|x|\le x_{0}, \qquad x_{0}>0$,}\\
C\, \sqrt{|x|+1}\,\text{K}_{\frac{1}{2}\,\sqrt{1-\lambda}}(\alpha\,(|x|+1)) &\text{if $|x|>x_{0}$.}
\end{cases}
\ee
where  $C$ is fixed by the requirement of continuity in $|x|=x_{0}$. Quantization amounts to the requirement
\be
\la{C.3}
\psi_{0}'(x_{0}^{-}) = \psi_{0}'(x_{0}^{+}),
\ee
and gives a transcendental relation between $\alpha$ and $\lambda$. Expanding at small $\lambda$, we find 
\be
\la{C.4}
\begin{split}
& \sqrt{-E_{0}(\lambda)} = \frac{2\,x_{0}+1}{4\,(x_{0}+1)^{2}}\,\lambda\,\bigg[
1+\frac{1}{2}\,\text{L}\,\lambda+\bigg(
\frac{1}{8}\,\text{L}^{2}+\frac{3\,x_{0}^{2}+10\,x_{0}+5}{8\,(x_{0}+1)^{2}}\,\text{L}\\
&
-\frac{896\, x_0^6+5988\,
   x_0^5+15735\, x_0^4+19620\, x_0^3+12510\, x_0^2+3960\, x_0+495}{1440\, \left(x_0+1\right)^4 \left(2
   x_0+1\right)^2}\bigg)\,\lambda^{2}\\
& +\bigg(
\frac{1}{48}\,\text{L}^{3}+\frac{3 x_0^2+14 x_0+7}{16 \left(x_0+1\right){}^2}\,\text{L}^{2}\\
&+\frac{
1744 x_0^6+12492 x_0^5+31725 x_0^4+38340 x_0^3+24030 x_0^2+7560 x_0+945}{2880
   \left(x_0+1\right){}^4 \left(2 x_0+1\right){}^2}\,\text{L}+\text{const}
\bigg)\,\lambda^{4}+\cdots
\bigg], \\
\text{L} &= \log\lambda+\gamma_{\rm E}+\log\left(\frac{2x_{0}+1}{2\,(x_{0}+1)}\right)
-\frac{8\,x_{0}^{3}+3\,x_{0}^{2}-6\,x_{0}-3}{6\,(x_{0}+1)^{2}(2\,x_{0}+1)}.
\end{split}
\ee
Of course, the expansion (\ref{C.4}) reduces to (\ref{2.3}) for $x_{0}=0$
and is consistent with (\ref{4.39}) and (\ref{4.40}) for generic $x_{0}$ -- 
the prefactor in (\ref{C.4})
is  $-\tfrac{1}{2}\,\Vb$. The leading logarithms are precisely the same as in (\ref{2.3}). We pushed further
the expansion (\ref{C.4}) identifying the rational coefficients of the higher leading logarithms and they 
continue to agree with (\ref{2.3}). This strongly supports the exponentiation (\ref{7.1}), where the prefactor 
depends on the global shape of $V$, while the exponential is universal in the sense that it 
depends only on the coefficient of the leading term 
of $V$
at infinity. Instead, the subleading logarithms depend on $x_{0}$ both through
the scale entering the definition of $\text L$ as well as in the 
remaining coefficient of the $\lambda^{n+2}\,\text{L}^{n}$ contributions.

\bibliography{QQbar-Biblio}
\bibliographystyle{JHEP}

\end{document}